\documentclass[useAMS,usenatbib,fleqn]{mnras}
\usepackage{multirow}
\usepackage{graphicx}
\usepackage{color,soul}
\usepackage{graphicx}
\usepackage{subfig}
\usepackage{epsfig}
\usepackage{epstopdf}
\usepackage{amssymb,amsmath}
\usepackage{aas_macros}
\usepackage{fixltx2e}
\usepackage[T1]{fontenc}

\title[Magnetic Fields and Circular Polarization]{
Black Hole Magnetic Fields and Their Imprint on Circular Polarization Images
}

\author[Ricarte et al.]{Angelo Ricarte$^{1,2}$, Richard Qiu$^{3,4}$, and Ramesh Narayan$^{1,2}$
\\
$^{1}$ Center for Astrophysics | Harvard \& Smithsonian, 60 Garden Street, Cambridge, MA 02138, USA \\
$^{2}$ Black Hole Initiative at Harvard University, 20 Garden Street, Cambridge, MA 02138, USA \\
$^3$ Department of Physics, Harvard University, 17 Oxford Street Cambridge, MA 02138, USA \\
$^4$ John A. Paulson School of Engineering and Applied Sciences, Harvard University, 29 Oxford Street, Cambridge, MA 02138, USA
}

\date{\today}

\begin{document}
\pagerange{\pageref{firstpage}--\pageref{lastpage}} \pubyear{2020}
\maketitle

\begin{abstract}
The circular polarization of black hole accretion flows can encode properties of the underlying magnetic field structure.  Using general relativistic magnetohydrodynamics (GRMHD) simulations, we study the imprint of magnetic field geometry on circular polarization images potentially observable by the Event Horizon Telescope (EHT).  We decompose images into the different mechanisms that generate circular polarization in these models, which are sensitive to both the line of sight direction and twist of the magnetic field.  In these models, a stable sign of the circular polarization over time, as observed for several sources, can be attributed to a stability of these properties.  We illustrate how different aspects of a generic helical magnetic field geometry become imprinted on a circular polarization image.  We also identify novel effects of light bending that affect the circular polarization image on event horizon scales.  One consequence is the sign flipping of successive photon rings in face-on systems, which if observable and uncorrupted by Faraday rotation, can directly encode the handedness of the approaching magnetic field.
\end{abstract}


\section{Introduction}
\label{sec:introduction}

Supermassive black holes (SMBHs) are thought to reside at the centres of all massive galaxies, where they sometimes shine as active galactic nuclei (AGN) and help regulate star formation via AGN feedback \citep[e.g.,][]{Kormendy&Ho2013}.  Detailed studies of their central engines are important to better understand their accretion and feedback processes.  By performing very long baseline interferometry (VLBI) at 1.3 millimetres, the Event Horizon Telescope (EHT) collaboration produced the first resolved images of the SMBH at the centre of M87, enabling novel studies of accretion physics on event horizon scales \citep{EHT1,EHT2,EHT3,EHT4,EHT5,EHT6,EHT7,EHT8}.  

The SMBH of M87, henceforth M87*, is believed to be fed by an advection dominated accretion flow (ADAF) of hot, tenuous plasma \citep{Ichimaru1977,Rees+1982,Narayan&Yi1994,Narayan&Yi1995,Abramowicz+1995,Reynolds+1996a,Yuan&Narayan2014}.  Its image is sensitive to properties intrinsic to the SMBH, including spin and accretion rate, as well as those of the plasma in the surrounding environment, such as the magnetic field configuration, temperature, and underlying electron distribution function.  Combining multi-wavelength constraints with the first resolved image of M87* enabled \citet{EHT5} to narrow the allowed parameter space of M87*, ultimately ruling out all non-spinning SMBH models.

Recently, the EHT Collaboration completed an analysis of the linear polarization of M87*, providing novel insights into its magnetic field structure in particular \citep{Goddi+2021,EHT7,EHT8}.  Synchrotron emission, which dominates the millimetre image, is initially generated perpendicular to the local magnetic field, such that its polarization carries with it an imprint of the field geometry \citep{Palumbo+2020}.  Then, as this polarization propagates, it is further modified by Faraday effects, important for depolarizing accretion flows down to observed levels and generating the observed rotation measure \citep{Ballantyne+2007,Moscibrodzka+2017,Jimenez-Rosales&Dexter2018,Ricarte+2020}.  Fully polarized radiative transport simulations on EHT scales have been developed in the past few decades, allowing us to link polarized images to the detailed physics of the underlying plasma and the space-time producing them \citep{Bromley+2001,Broderick&Loeb2006,Broderick&McKinney2010,Porth+2011,Shcherbakov+2012,Dexter2016,Moscibrodzka&Gammie2018}.  This enabled \citep{EHT8} to discriminate between two major classes of accretion disk:  a ``Magnetically Arrested Disk'' (MAD) and ``Standard and Normal Evolution'' (SANE).  MAD accretion disks have magnetic fields strong enough to affect the disk dynamics and exhibit stronger poloidal (or non-toroidal) magnetic field components \citep{Bisnovatyi-Kogan&Ruzmaikin1974,Igumenshchev+2003,Narayan+2003,Chael+2019}.  Meanwhile, the weaker magnetic fields of a SANE disk are sheared out by the motion of the plasma into a mostly toroidal configuration \citep{Narayan+2012,Sadowski+2013,Ryan+2018}.  The fractional linear polarization of M87*, the upper limit on its circular polarization, and most importantly the ``twisty pattern'' of its spatially resolved linear polarization map favour ``MAD'' models of M87* \citep{EHT8}.

In addition to these recent studies of M87*, there exist unresolved polarimetric measurements of Sgr A* and other low-luminosity AGN \citep{Plambeck+2014,Kim+2019}.  In particular, linear polarization measurements of Sgr A* in the millimetre span two decades, motivating further studies of time variable polarimetry \citep{Aitken2000,Bower+2003,Marrone2006,Munoz+2012,Bower+2018}.  Moreover, although there is only an upper limit on the total circular polarization of M87* so far \citep[$\lesssim 0.8$ per cent][]{Goddi+2021}, the EHT Collaboration plans to one day produce images of circular polarization in addition to linear.  Circularly polarized images of SMBH accretion flows are less well studied than their linearly polarized counterparts.  For several observed sources, a persistent magnitude and sign of total circular polarization has been interpreted as stability in something inherent to the accretion flow such as its magnetic field geometry, but the details are poorly understood \citep{Wardle&Homan2001,Beckert&Falcke2002,Ruszkowski&Begelman2002,Ensslin+2003}.  One recent study of MAD accretion disks highlights the importance of both Faraday rotation and conversion and points out a sign inversion in the photon ring \citep{Moscibrodzka+2021}.

There are two mechanisms that can generate circular polarization in these systems:  intrinsic emission of circular polarization, and Faraday conversion of linear polarization \citep[e.g.,][]{Wardle&Homan2003}.  Synchrotron emission is the dominant emission mechanism in the millimetre, and it has an intrinsic linear polarization fraction of about 70 per cent \citep[e.g.,][]{Pandya+2016}.  In contrast, its emitted circular polarization fraction is only around 1 per cent, but this may be significant enough to contribute to the modest circular polarization fractions that have been measured for low-luminosity AGN.  The sign of intrinsically emitted circular polarization directly encodes the direction of the magnetic field with respect to the photon wavevector.  Meanwhile, Faraday conversion exchanges linear and circular polarization, with an efficiency depending on the temperature, density, and magnetic field of the plasma through which the light is travelling \citep[e.g.,][]{Kennett&Melrose1998}.  While the intensity of intrinsic emission increases with temperature, Faraday conversion is more efficient at lower (sub-relativistic) temperatures \citep{Jones&Odell1977}.  Both the intrinsic circular polarization fraction and the strength of Faraday conversion are sensitive to the composition of the underlying plasma.  This allows circular polarization measurements to provide some of the most direct constraints on plasma composition \citep{Wardle+1998,Anantua+2020,Emami+2021}.

Interestingly, Faraday conversion has no effect in a plasma with a uni-directional magnetic field without an additional mechanism to rotate the polarization plane.  A synergy between Faraday conversion and Faraday rotation is one mechanism to accomplish this, as discussed in-depth with GRMHD models in \citet{Tsunetoe+2020a,Tsunetoe+2020b}.  Alternatively, a line of sight twist in the magnetic field can also effectively rotate the polarization, by instead rotating the basis in which the polarization states are defined \citep{Hodge1982,Wardle&Homan2001}.  This phenomenon is believed to set the circular polarization of jets on larger scales \citep[e.g.,][]{Gabuzda+2008}.  As we shall show, a more complicated magnetic field structure, combined with light bending effects, produces a rich circular polarization image on event horizon scales.

In this work, we study in detail the mechanisms which generate circular polarization in SMBH accretion flows using two GRMHD simulations, described in \S\ref{sec:methodology}.  To better understand the connection between circular polarization and magnetic field, we first study the structure of the magnetic field in these models in \S\ref{sec:magnetic_field_decompositions}.  Then, in \S\ref{sec:circular_polarization}, we study each of the mechanisms for generating circular polarization in turn.  We discuss the more general implications of this study in \S\ref{sec:discussion}, and summarise our results in \S\ref{sec:conclusion}.

\section{Methodology}
\label{sec:methodology}

We use as our starting point two GRMHD simulations of M87* that are included in \citet{EHT5,EHT8}, one ``Magnetically Arrested Disk'' (MAD) and one ``Standard and Normal Evolution'' (SANE), both run using {\sc harm} \citep{Gammie+2003}.  Although a SANE model is disfavoured for M87* based on linear polarization \citep{EHT8}, it remains instructive to explore how such a system differs in its circularly polarized image.  Both models have dimensionless spin parameter $a_\bullet=0.94$ and are run using a variation of Kerr-Schild coordinates (``Funky Modified Kerr-Schild,'' or FMKS) that concentrates resolution at the mid-plane.  These simulations are seeded with a weak dipolar magnetic field parallel to both the spin of the BH and the angular momentum of the accretion disk.  The MAD simulation is performed using a $384\times 192\times 192$ grid with a maximum radius of $10^3 \ GM_\bullet/c^2$, while the SANE simulation uses a $288\times 128\times 128$ grid and a maximum radius of $50 \ GM_\bullet/c^2$.  (Throughout this work, $G$ is the gravitational constant, $M_\bullet$ is the SMBH mass, and $c$ is the speed of light.) As in \citet{Ricarte+2020}, we only evolve the radiative transfer equations within a radius of $20 \ GM_\bullet/c^2$, inside of which we find the simulations exhibit inflow equilibrium.  In Appendix \ref{sec:pixel_traces}, we demonstrate that this region encloses the relevant plasma for producing circular polarization in these models.  Both models are evolved to $10^4 \ GM_\bullet/c^3$, and we focus our presentation on the images of their final snapshots.

To perform the General Relativistic Ray Tracing (GRRT), we use {\sc ipole}, a ray-tracing code which first solves the null geodesic equation from the camera through the source, then integrates forward the fully polarized radiative transfer equation \citep{Moscibrodzka&Gammie2018}.  Unless otherwise noted, images are created at a frequency of 228 GHz, with a field of view of 160 $\mu$as, and an angular resolution of 0.5 $\mu$as.  As in many previous works, the plasma is modelled as two-temperature, where the ratio of the ion temperature to the electron temperature is a function of the plasma $\beta$ parameter $\beta=p_\mathrm{gas}/p_\mathrm{mag}$ via

\begin{equation}
    \frac{T_i}{T_e} = R_\mathrm{high}\frac{\beta^2}{1+\beta^2} + \frac{1}{1+\beta^2}\;,
\end{equation}

\noindent where $p_\mathrm{gas}$ and $p_\mathrm{mag}$ are the gas and electron pressures respectively \citep{Moscibrodzka+2016}.  As a consequence, $T_i = T_e$ in the highly magnetised funnel region, while $T_i/T_e \to R_\mathrm{high}$ in the gas pressure supported mid-plane.  For details of the radiative transfer coefficients, we refer readers to \citet{Moscibrodzka&Gammie2018}, and the modification to the Faraday rotation coefficient $\rho_V$ in \citet{Ricarte+2020}.  

In the two models considered in this work, we fix $R_\mathrm{high}=20$, a modest value motivated by recent studies of plasma heating (Mizuno et al.~in prep.).  For images with $R_\mathrm{high}=160$, we comment that the signals we discuss in the paper remain largely unchanged, but there is additional ``noise'' and depolarization due to the increased Faraday rotation.   This is because larger values of $R_\mathrm{high}$ decrease the electron temperature in the gas pressure supported mid-plane, by construction, making Faraday effects more efficient.  Meanwhile, for $R_\mathrm{high}=1$, the MAD image is largely unchanged, but the SANE image exhibits a stronger photon ring feature discussed in \S\ref{sec:magnetic_twist_faceon}, which is suppressed for $R_\mathrm{high}=20$ due to Faraday rotation.

Since GRMHD simulations are scale free, the SMBH mass, distance, and accretion rate are set during the GRRT step.  To set the spatial and temporal scales, we adopt a mass of $6.2 \times 10^9 \ \mathrm{M}_\odot$ and a distance of 16.9 Mpc, consistent with stellar dynamics and shadow size measurements of M87* \citep{Gebhardt+2011,EHT6}.  Then, we adopt the accretion rate scalings that are fit in \citet{EHT5} to reproduce an average flux of 0.5 Jy at 230 GHz.  Such scaling is possible because ideal GRMHD simulations are invariant under the transformation $\rho \mapsto \mathcal{M} \rho$ and $\vec{B} \mapsto \mathcal{M}^{1/2} \vec{B}$, where $\rho$ is the mass density, $\vec{B}$ is the magnetic field vector, and $\mathcal{M}$ is a scalar.

\section{Magnetic Field Structure}
\label{sec:magnetic_field_decompositions}

\subsection{How Circular Polarization Encodes the Magnetic Field}

\begin{figure*}
\begin{minipage}[c][]{0.5\textwidth}
\subfloat{\includegraphics[width=\textwidth]{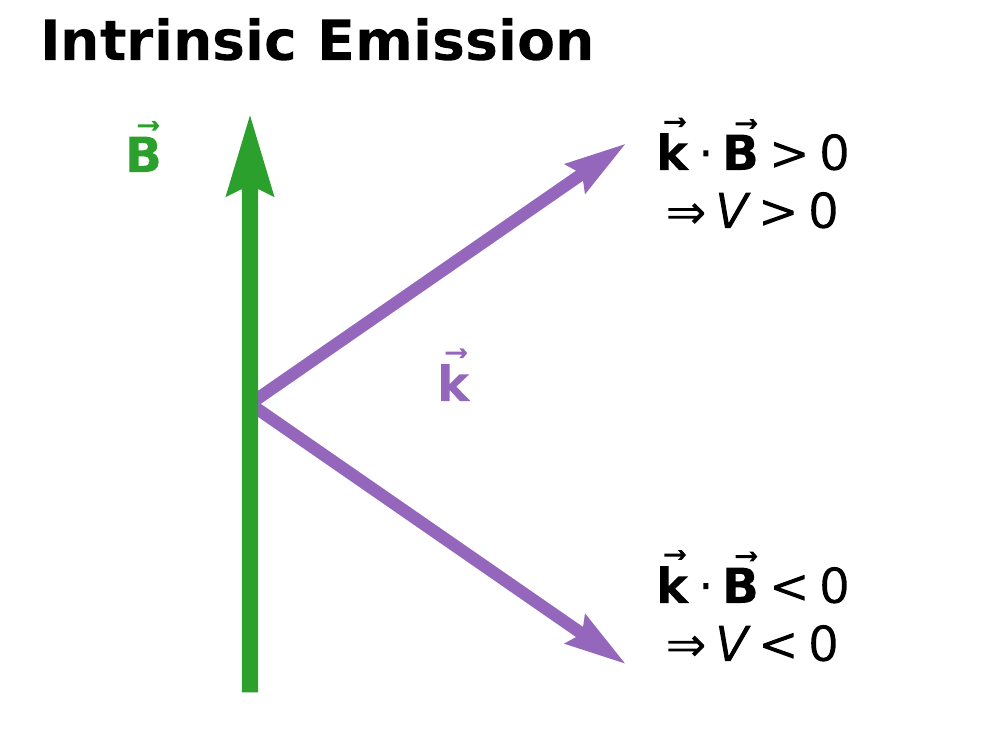}}
\hrule
\vfill
\subfloat{\includegraphics[width=\textwidth]{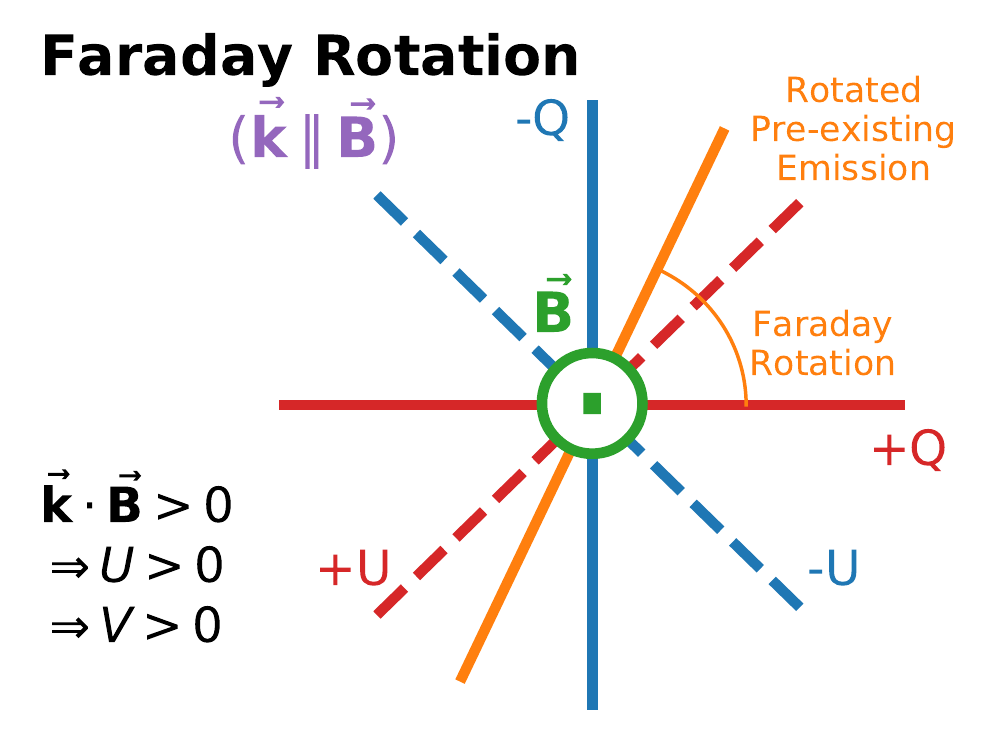}}
\end{minipage}%
\vline
\begin{minipage}[c][]{0.5\textwidth}
\subfloat{\includegraphics[width=\textwidth]{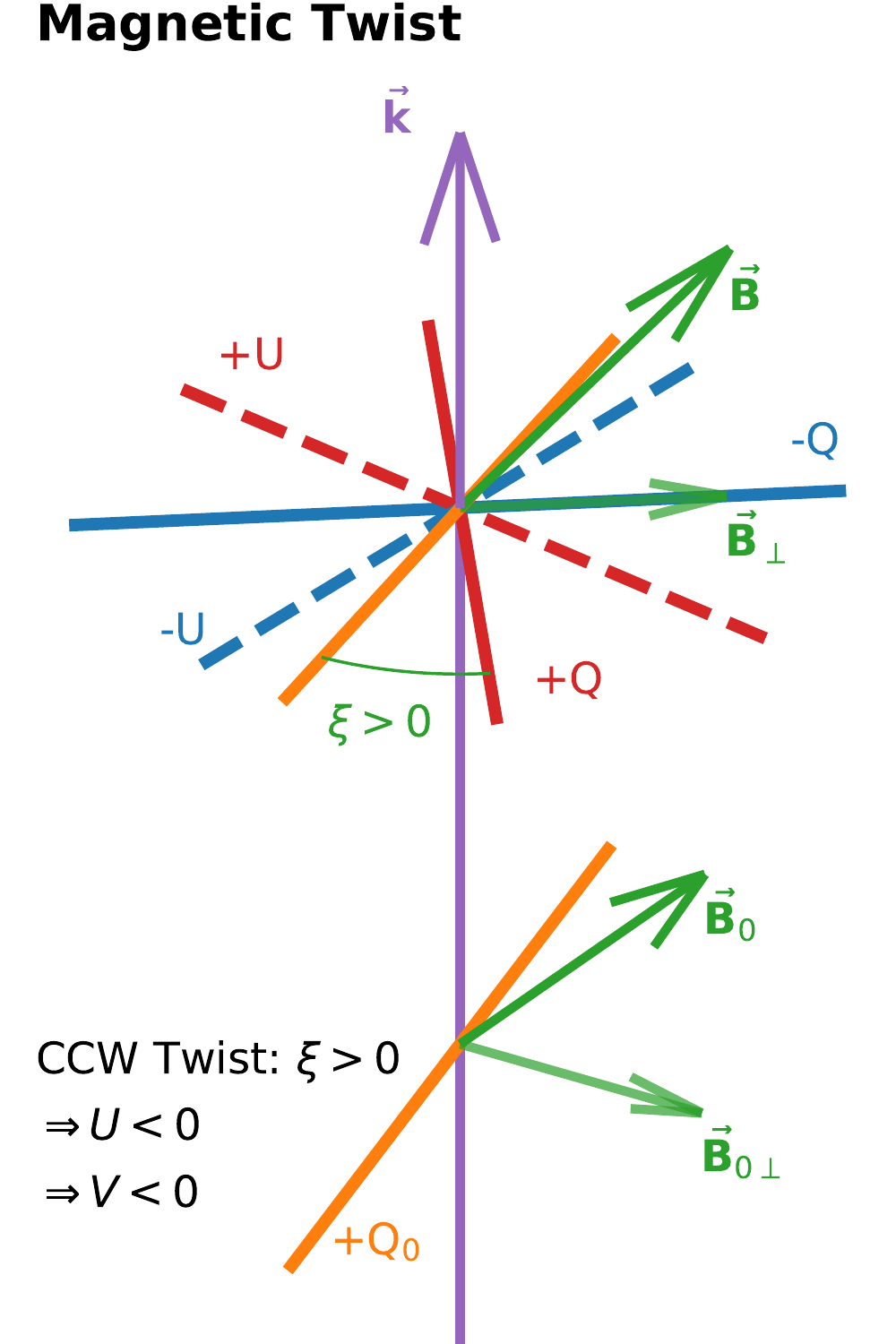}}
\end{minipage}%
\caption{Schematics illustrating the mechanisms by which circular polarization can be generated and the resultant signs of Stokes $V$.  The sign of intrinsically emitted circular polarization directly reflects the magnetic field direction with respect to the photon wavevector (Top Left). Intrinsically emitted circular polarization is positive if the magnetic field is pointed towards the observer, and negative if it is pointed away.  Faraday conversion only operates on Stokes $U$, and thus Stokes $V$ inherits the local sign of Stokes $U$ from mediating phenomena.  Circular polarization generated through Faraday conversion mediated by a small amount of Faraday rotation (Bottom Left) again encodes the line of sight magnetic field direction such that $\vec{k}\cdot\vec{B}>0$ leads to $V>0$.  If the magnetic field twists along the line of sight (Top and Bottom Right), then circular polarization generated through Faraday conversion acquires a sign opposite to the sign of the twist $\xi$, since it is the basis in which the Stokes parameters are defined that is twisted, rather than the emission itself. Therefore, a counter-clockwise (CCW) twist of $\xi>0$ generates $V<0$.}\label{fig:schematics_many}
\end{figure*}

\begin{figure}
  \centering
  \includegraphics[width=0.4\textwidth]{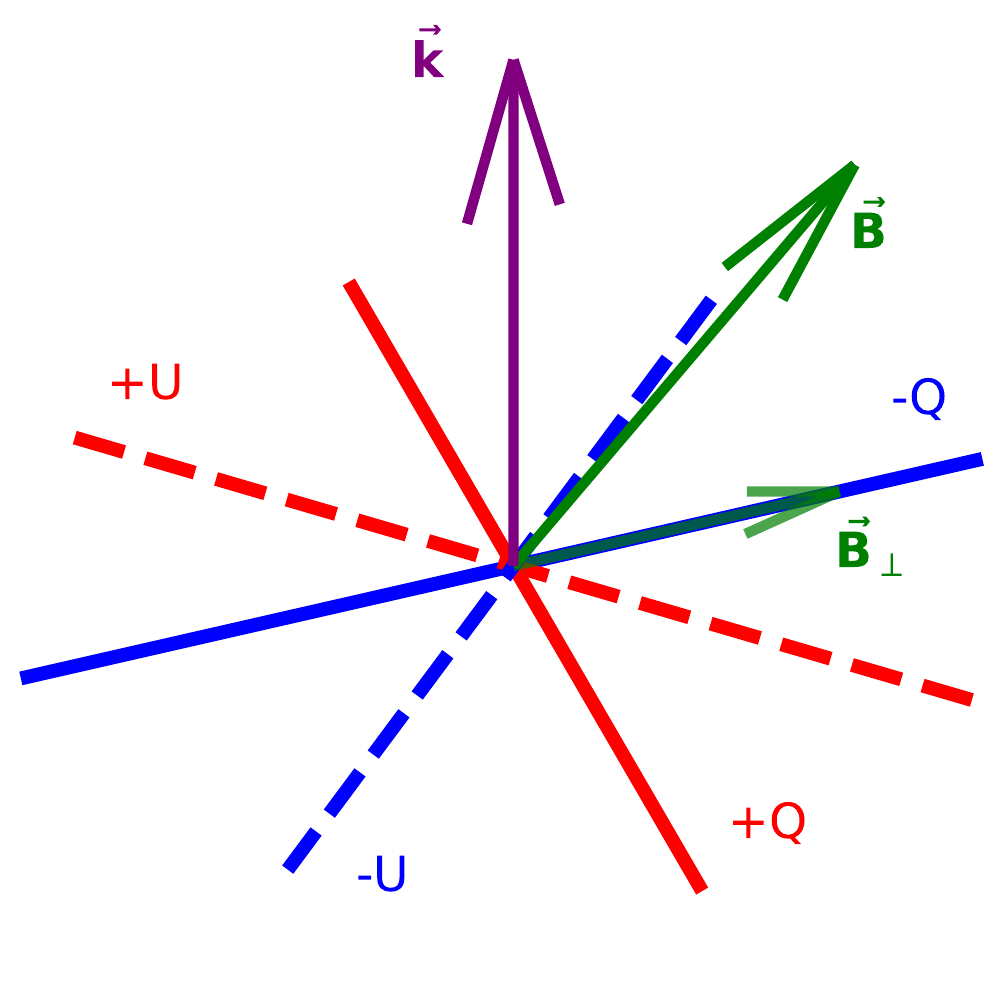}
  \caption{Schematic showing the definitions of the Stokes parameters in the local tetrad frame, defined by the photon wavevector $\vec{k}$ and the magnetic field vector $\vec{B}$.  Synchrotron emission produces linearly polarized emission in the $+Q$ orientation, perpendicular to both the photon wavevector and the magnetic field.  \label{fig:plasma_tetrad}}
\end{figure}

There are three pathways to generate circular polarization, each of which reflects a different aspect of the magnetic field:  intrinsic emission, Faraday conversion mediated by Faraday rotation, and Faraday conversion mediated by a line of sight twist in the magnetic field \citep[e.g.,][]{Wardle&Homan2003}.  These processes are illustrated in Figure \ref{fig:schematics_many}.  

First, synchrotron emission intrinsically produces a small amount of circular polarization.  The sign of this circular polarization directly encodes the line of sight direction of the magnetic field $\vec{B}$ with respect to the photon wavevector $\vec{k}$. The emitted radiation has positive $V$ if the two vectors are aligned ($\vec{k}\cdot\vec{B}>0$) and negative $V$ if they are anti-aligned ($\vec{k}\cdot\vec{B}<0$).

At the same time, synchrotron emission produces a large linear polarization fraction in these models, oriented as illustrated in Figure \ref{fig:plasma_tetrad}.  In the local plasma frame, linear polarization is initialised perpendicular to both the magnetic field and the photon wavevector, which is locally defined as the $+Q$ direction (parallel to the electric field). As this linear polarization propagates along $\vec{k}$, some of it can be converted to circular via Faraday conversion.  However, Faraday conversion only operates on Stokes $U$ in the local plasma frame, a component neither parallel nor perpendicular to $\vec{k}\times\vec{B}$.  If $U>0$, the resulting circular polarization has $V>0$, and vice versa.  Since the intrinsic synchrotron emission has only $+Q$ (i.e., $U=0$), something has to recast a part of the emitted $Q$ into $U$ as the radiation propagates through the plasma for Faraday conversion to occur. This can happen in two ways, corresponding to two pathways through which Faraday conversion can operate.

Faraday rotation can change the plane of polarization as the radiation propagates by directly exchanging $Q$ and $U$.  The sense of rotation operates via the ``right hand rule'' on the direction of the line-of-sight component of the magnetic field. For small amounts of Faraday rotation, if $\vec{k}\cdot\vec{B}>0$, Faraday rotation converts $+Q$ partially to $+U$, and Faraday conversion then creates circular polarization with $+V$.  Similarly, $\vec{k}\cdot\vec{B}<0$ leads to $-V$.

The other way in which Faraday conversion can operate is if the component of the magnetic field perpendicular to $\vec{k}$ rotates because of a line of sight twist in the magnetic field.  Since the basis in which the Stokes parameters are defined is tied to the local magnetic field, if the field twists along the line of sight, radiation that was initially emitted entirely as Stokes $+Q$ can be ``recast'' as partially Stokes $U$, with a sign depending on the direction and degree of the twist.  In this pathway, since it is the polarization basis rather than the radiation itself that is rotating, a small positive (counter-clockwise, CCW) twist in the magnetic field produces $-U$ and consequently through Faraday conversion $-V$. 

We caution that the above discussion on the sign of $V$ resulting from Faraday conversion applies only for small relative rotations of the plane of polarization ($<90^\circ$).  For larger rotations, the sign of $U$ generated by either Faraday rotation and/or magnetic field twist flips each time the relative polarization plane is rotated by an additional $<90^\circ$.  In practice, the large Faraday rotation depth in some of these models \citep{Moscibrodzka+2017,Jimenez-Rosales&Dexter2018,Ricarte+2020} can also depolarize circular polarization in their images, by essentially randomising the sign of $U$ which would be converted into $V$, as we will further discuss in \S\ref{sec:faraday_conversion}.

\subsection{Generic Magnetic Field Properties}
\label{sec:magnetic_field_generic}

\begin{figure}
  \centering
  \includegraphics[width=0.5\textwidth]{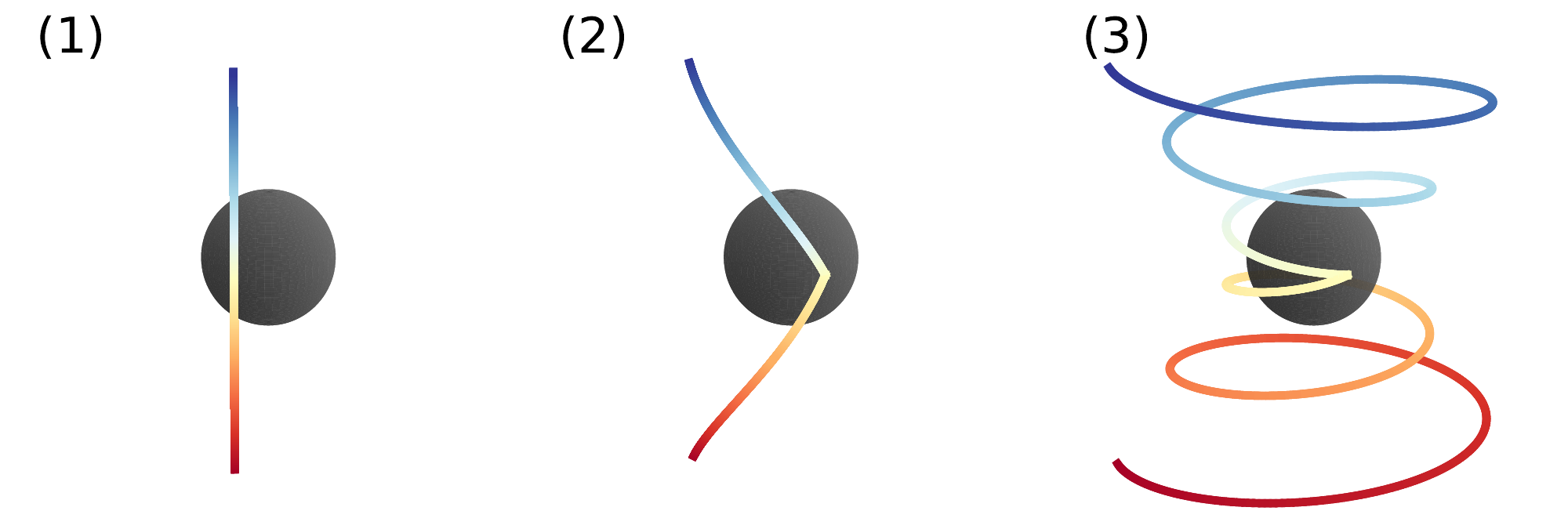}
  \caption{Cartoon illustrating the formation of a twisted magnetic field.  A vertically oriented field line is twisted by frame dragging and flux freezing.  Notice the opposite handedness of the twist generated above and below the mid-plane.  The colours of these curves simply denote relative position along the field lines. \label{fig:cartoon_field_line}}
\end{figure}

Since the mechanisms to generate circular polarization depend on the details of the magnetic field, let us first explore its three-dimensional geometry.  Assuming an initial dipolar magnetic field structure, frame dragging and flux freezing produce a generic helical magnetic field pattern, as illustrated in Figure \ref{fig:cartoon_field_line} \citep[e.g.,][]{Semenov+2004}.  Notice that the handedness of these helices flips across the mid-plane, resulting in a sign flip in the radial and tangential components of the magnetic field.

\begin{figure}
  \centering
  \includegraphics[width=0.5\textwidth]{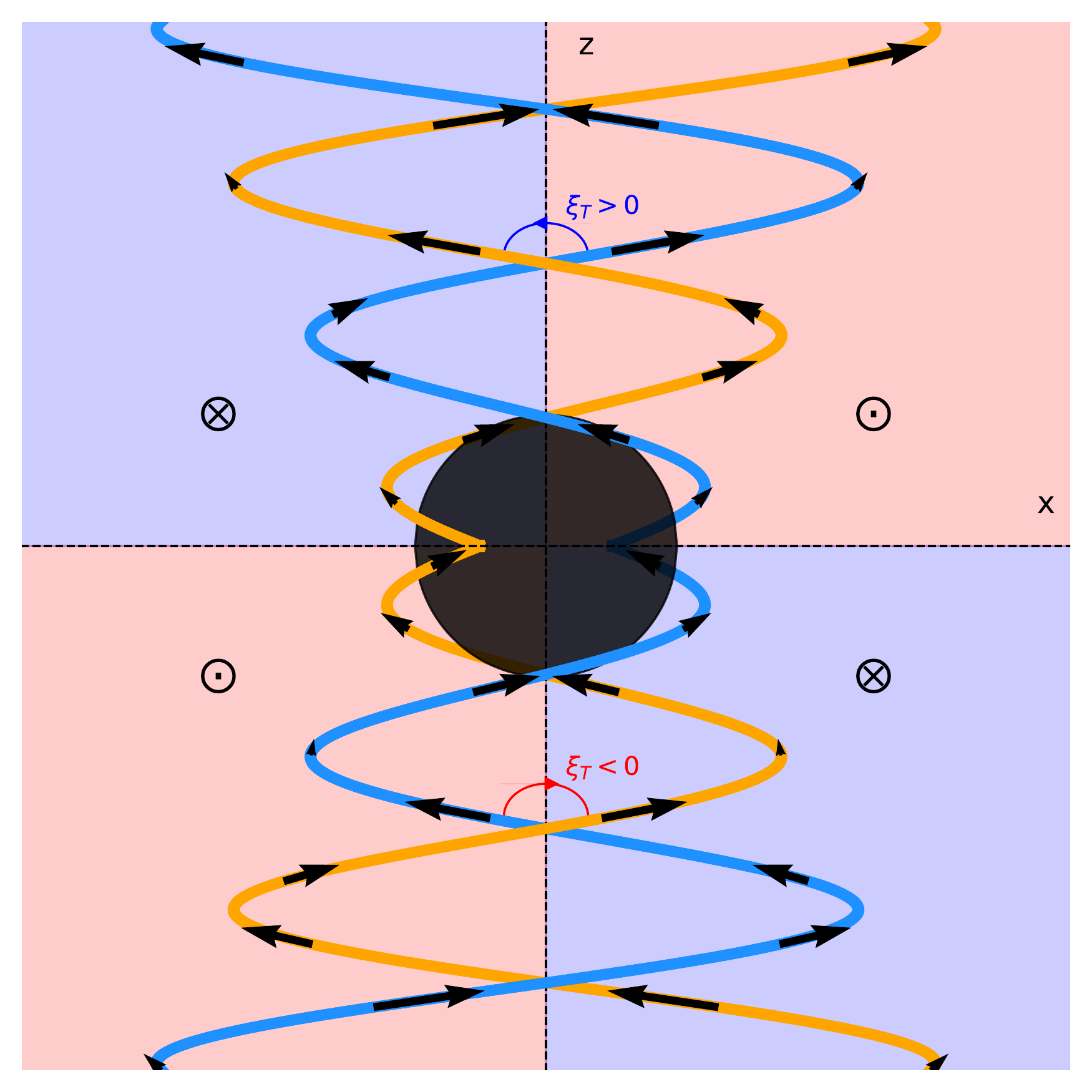}
  \caption{Schematic illustrating two field lines anchored at opposite sides of the disk mid-plane.  The $\bigodot$ and $\bigoplus$ symbols indicate whether the magnetic field lines move out of or into the page.  Note the four quadrants generated by the sign flip in the tangential and radial components across the mid-plane, as illustrated in Figure \ref{fig:cartoon_field_line}.  At two locations chosen at opposite sides of the mid-plane, the angle $\xi_T$ illustrates the magnetic twist direction between the back and front sides, what we term the ``transverse twist.''  Since the top and bottom helices have opposite handedness, $\xi_T$ flips sign across the mid-plane.  \label{fig:cartoon_transverse_twist}}
\end{figure}

The outcome of this process is the structure illustrated in Figure \ref{fig:cartoon_transverse_twist}, where blue and orange curves represent magnetic field lines anchored on opposite sides of the BH in the mid-plane.  We have assumed that the BH angular momentum, the disk angular momentum, and the direction of the dipolar magnetic field are all aligned with what we illustrate as the $z$ axis.  Applying Ampere's law, this configuration corresponds to an inward current towards the BH.  We comment that for a small sample of AGN with constrained jet magnetic field geometries, there does appear to be a preference for inward currents \citep{Gabuzda2018b}, which matches expectations from the cosmic battery mechanism \citep{Contopoulos&Kazanas1998,Koutsantoniou&Contopoulos2014}.  For a dipolar field initialised in the positive $z$ direction, the field lines head out of the page in red regions marked with the $\bigodot$ symbol, and into the page in blue regions marked with the $\bigoplus$ symbol.  Two sign flips occur in the line of sight magnetic field direction, across both the $x$ and $z$ axes in this cartoon, which is also pointed out by \citet{Tsunetoe+2020b}.  This pattern, which we refer to as the ``four quadrants,'' reverses if the dipolar field is instead initialised in the negative $z$ direction.  The sign flip across the $z$ axis is just a viewing angle effect due to rotation, while the sign flip across the mid-plane is physical, generated as described in Figure \ref{fig:cartoon_field_line}.  The mid-plane sign flip implies the existence of an equatorial current sheet, which would be subject to plasmoid instabilities and magnetic reconnection in higher resolution simulations that include resistivity \citep{Ripperda+2020}.

\subsection{Magnetic Twist}
\label{sec:magnetic_twist}

In the absence of physical effects to exchange Stokes $Q$ and $U$, the line of sight twist in the magnetic field determines the sign of circular polarization generated through Faraday conversion \citep{Gabuzda+2008}.  We write the magnetic twist along a line of sight as $\xi$, where a right-handed (counter-clockwise) twist has positive $\xi$, and a left-handed (clockwise) twist has negative $\xi$.\footnote{To be more explicit, the magnetic twist along the photon trajectory between points A and B is the angle, in the plane perpendicular to the photon wavevector at point B, between the local magnetic field vector and the vector resulting from parallel transport of the magnetic field vector from point A to point B.}  Throughout this work, we always consider twist as defined in the direction of the photon's propagation, through the source to the observer.  

During our explorations, we found that circular polarization generated via conversion mediated by magnetic twist could exhibit two different signs when viewing the same side of the disk tilted at different angles.  This is because there are two different relevant twists, depending on one's viewing inclination, which we term the ``transverse'' twist (for edge-on viewing angles) and ``vertical'' twist (for pole-on viewing angles).  By examining Figure \ref{fig:cartoon_transverse_twist}, we can determine the sign of what we term the ``transverse twist'' of these helices, $\xi_T$:  the relative twist between the far- and near-side magnetic field lines for this edge-on inclination.  These are illustrated by the blue and red arcs in this figure.  If the angular momentum of the system is oriented in the $+z$ direction, then $\xi_T>0$ above the mid-plane and $\xi_T<0$ below the mid-plane.  Note that the magnetic twist is invariant to the sign of the magnetic field; it would be the same if the arrows denoting the direction of the field were reversed.

\begin{figure}
  \centering
  \includegraphics[width=0.5\textwidth]{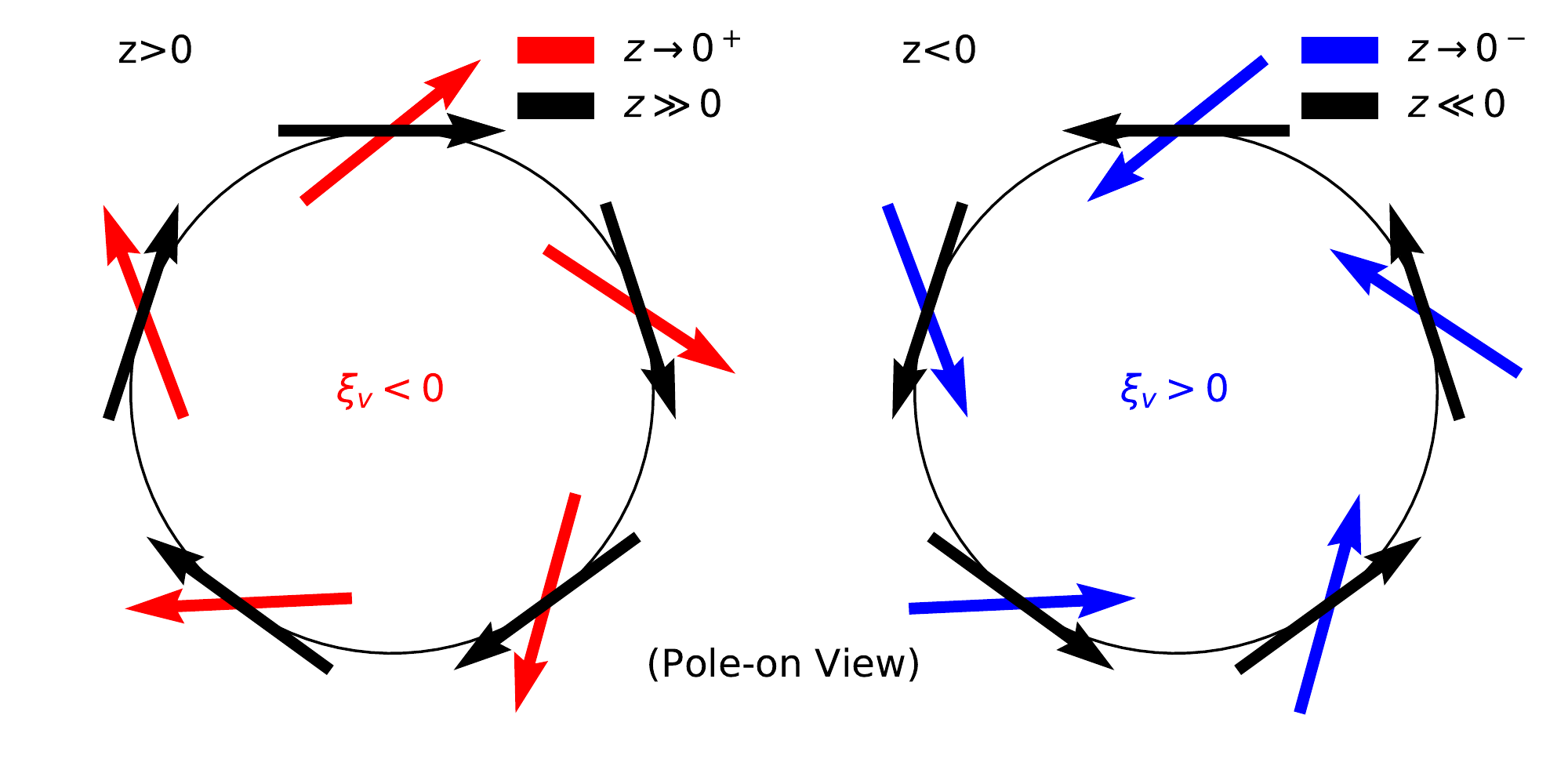}
  \caption{Schematic illustrating vertical twist, $\xi_V$, for a pole-on viewing angle.  Here, the angular momentum of the BH is in the positive $z$ direction, out of the page.  Because parabolic jets radially expand more rapidly near the base than at large distances, the radial component of a magnetic field threading such a structure grows weaker relative to the tangential component as $|z|$ increases.  Interestingly, comparing $\xi_V$ to $\xi_T$ in Figure \ref{fig:cartoon_transverse_twist}, the vertical twist in a given physical region exhibits the opposite sign as its transverse twist.  \label{fig:cartoon_vertical_twist}}
\end{figure}

We distinguish the ``vertical twist,'' $\xi_V$, from the ``transverse twist,'' $\xi_T$, as the twist of this structure when viewed from pole-on inclination angles.  The vertical twist is more relevant for M87* than its transverse twist, since we view the system at approximately $20^\circ$ inclination \citep{Walker+2018}.  The vertical twist direction can be deduced from generic parabolic jet structures, which widen more rapidly at the base than at larger distances \citep[e.g.,][]{Asada&Nakamura2012,Chatterjee+2019}.  This implies that as $|z|$ increases, the radial component of a magnetic field threading such a structure weakens relative to its tangential component.  In the funnel, it is actually the vertical field which dominates, which would go into or out of the page, but we are concerned only with the part of the magnetic field perpendicular to the propagation direction.  

From this, we can infer the sign of the vertical twist, as illustrated in Figure \ref{fig:cartoon_vertical_twist}.  In the $z>0$ region, $\xi_T > 0$ but $\xi_V < 0$, while the opposite is true for the $z<0$ region.  This implies that for generic accretion flows near the event horizon, {\bf vertical and transverse twist have opposite sign in the same region}.  When viewing the same side of the disk edge-on, the transverse twist is more important for determining circular polarization, while for pole-on inclination angles, the vertical twist is more important.  We confirm the sense of vertical twist in our GRMHD models in Appendix \ref{sec:app_vertical_twist}. 

\subsection{Magnetic Fields of Our GRMHD Models}
\label{sec:magnetic_fields_grmhd}

\begin{figure*}
  \centering
  \includegraphics[width=\textwidth]{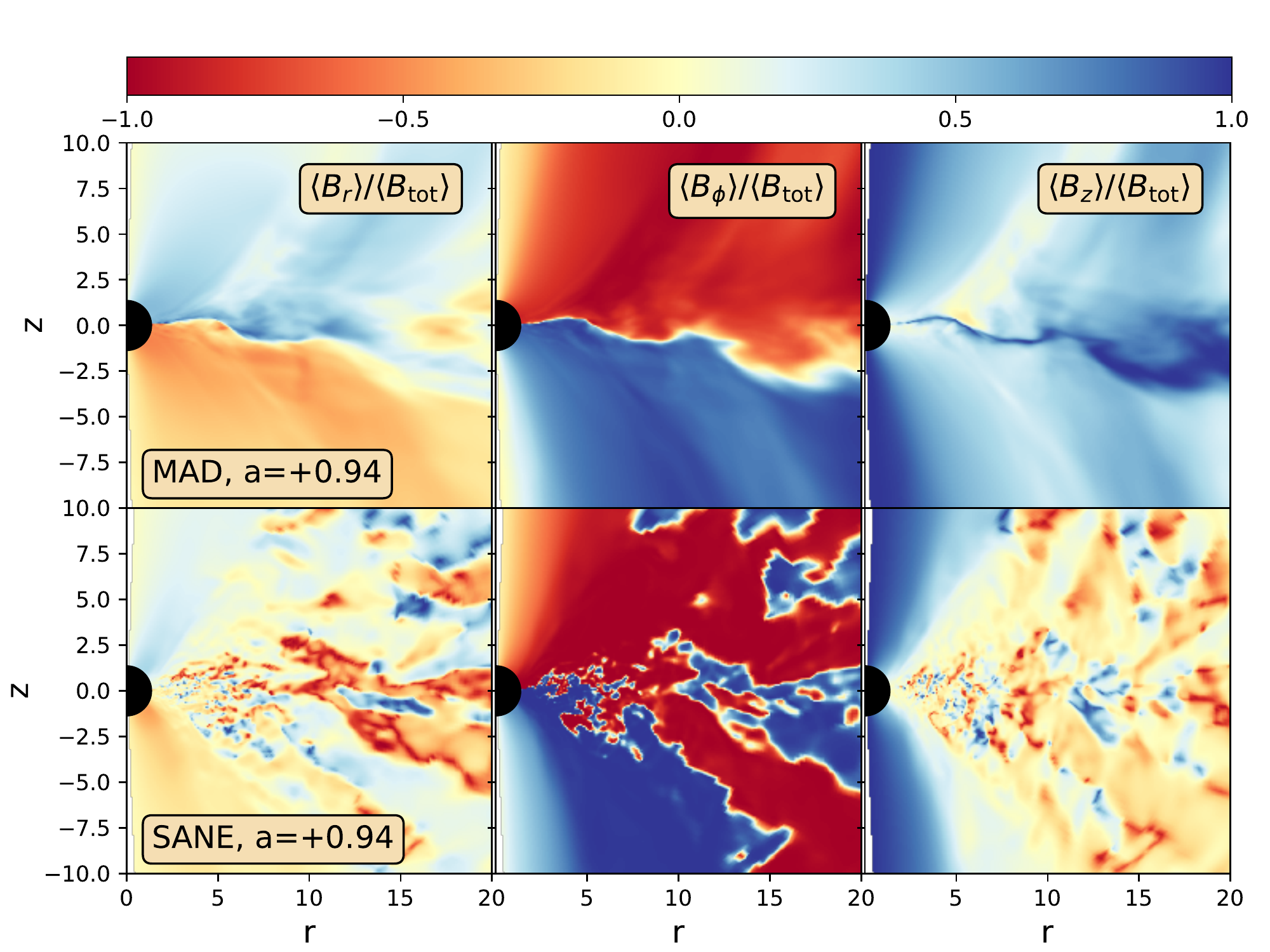}
  \caption{Azimuthally averaged magnetic field decompositions in the coordinate frame for the two models explored in this study during their final snapshots.  Notice (i) the dominance of the azimuthal field component in these particular models, (ii) the sign flip in the azimuthal and radial components across the mid-plane, and (iii) the greater disorder in the SANE model compared to the MAD.  \label{fig:b_decomposition}}
\end{figure*}

Next, we examine the magnetic field structure of our two GRMHD models to learn how they may depart from the simple pictures described in the previous section.  In Figure \ref{fig:b_decomposition}, we plot the magnetic fields in the coordinate frame in Cartesian Kerr-Schild coordinates. These are decomposed into their radial, azimuthal, and vertical components, then azimuthally averaged during the final snapshot, time $t=10000 \ GM_\bullet/c^3$.  These decompositions are normalised to the local magnetic field strength, such that $\langle B_\mathrm{tot} \rangle^2 = \langle B_r \rangle^2 + \langle B_\phi \rangle^2 + \langle B_z \rangle^2$.  There are three interesting phenomena to notice in this figure.  First, these two models generally have stronger azimuthal fields than radial fields, which need not generally be true.  The azimuthal component dominates everywhere except in the funnel, directly above and below the BH.  Second, notice the sign flip in the radial and tangential field components as discussed in the previous section.  Third, the SANE model is more turbulent and disordered than the MAD model, complicating the simple picture drawn in Figure \ref{fig:cartoon_transverse_twist}.  This turbulence will appear as noise in its circular polarization images.

\begin{figure*}
  \centering
  \begin{tabular}{cc}
  \includegraphics[width=0.5\textwidth]{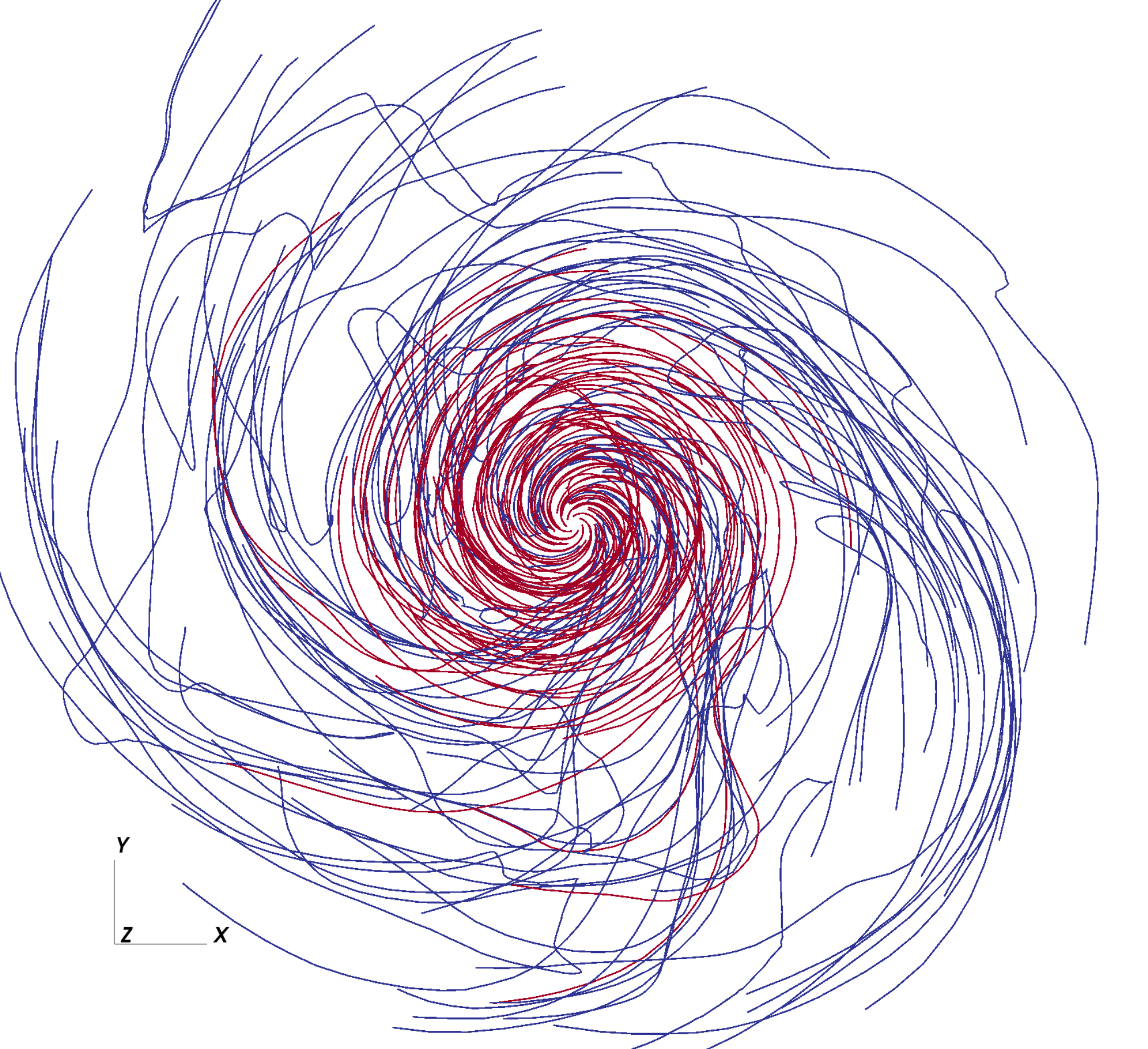} & 
  \includegraphics[width=0.5\textwidth]{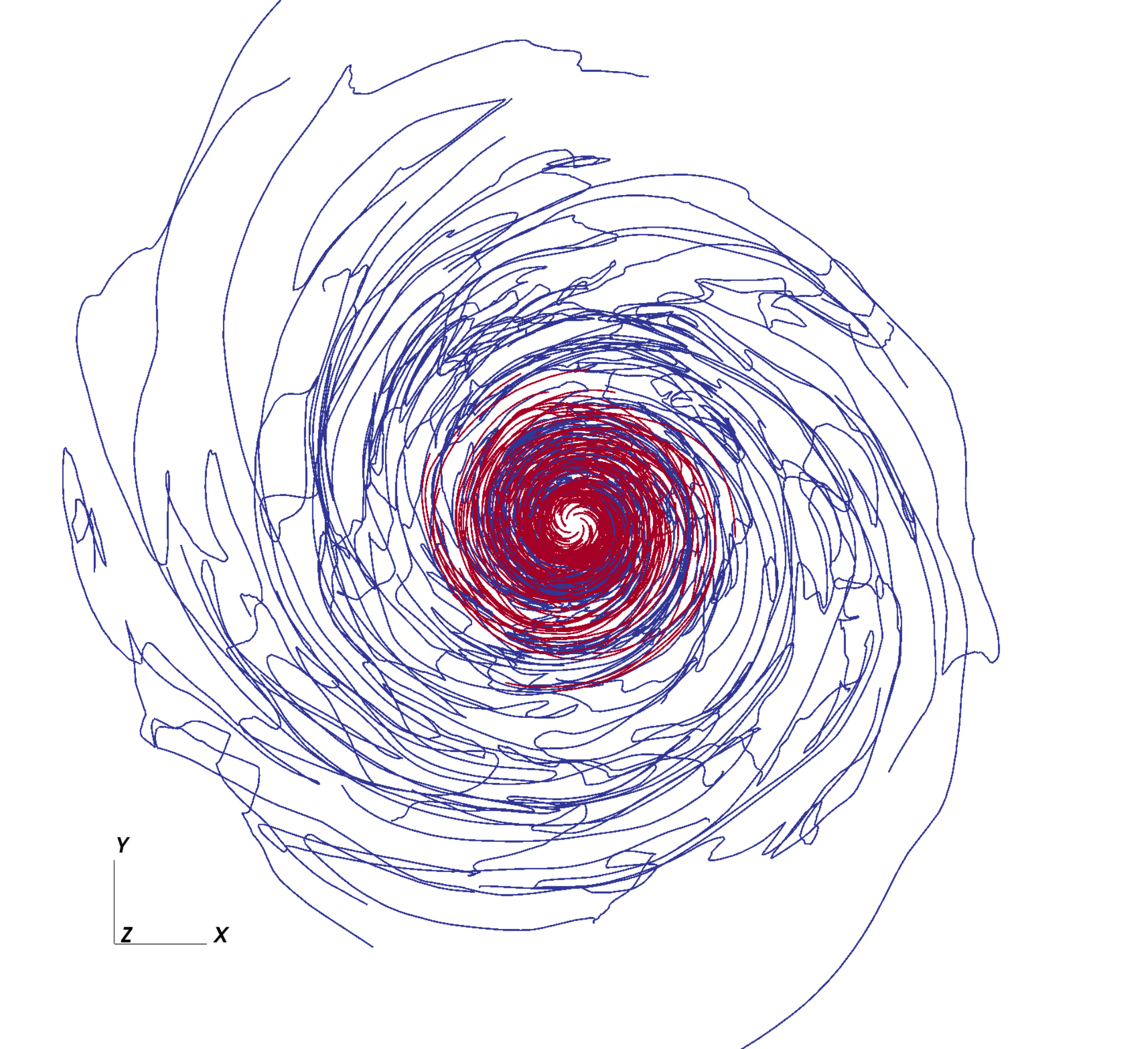} \\
  \includegraphics[width=0.5\textwidth]{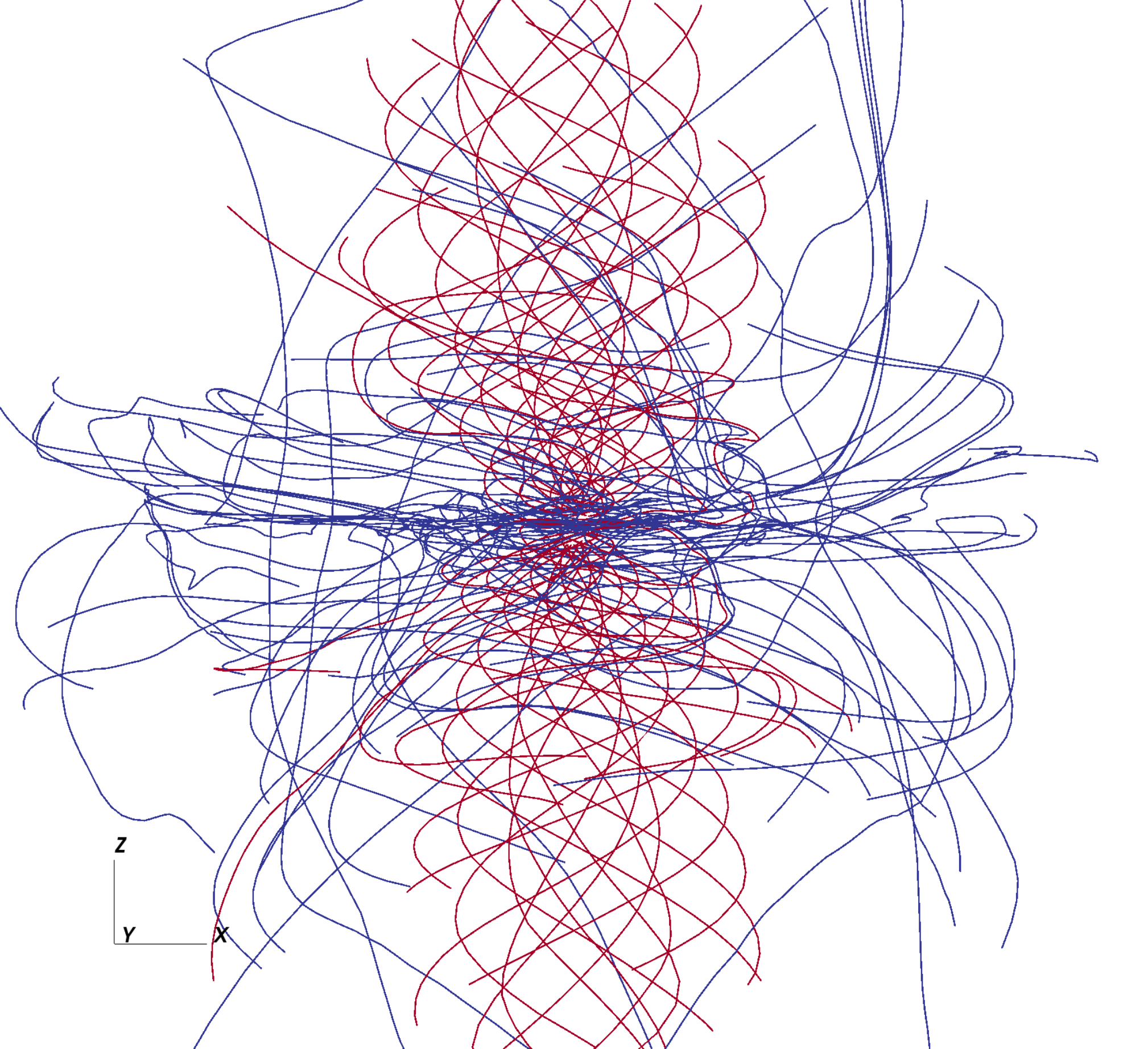} & 
  \includegraphics[width=0.5\textwidth]{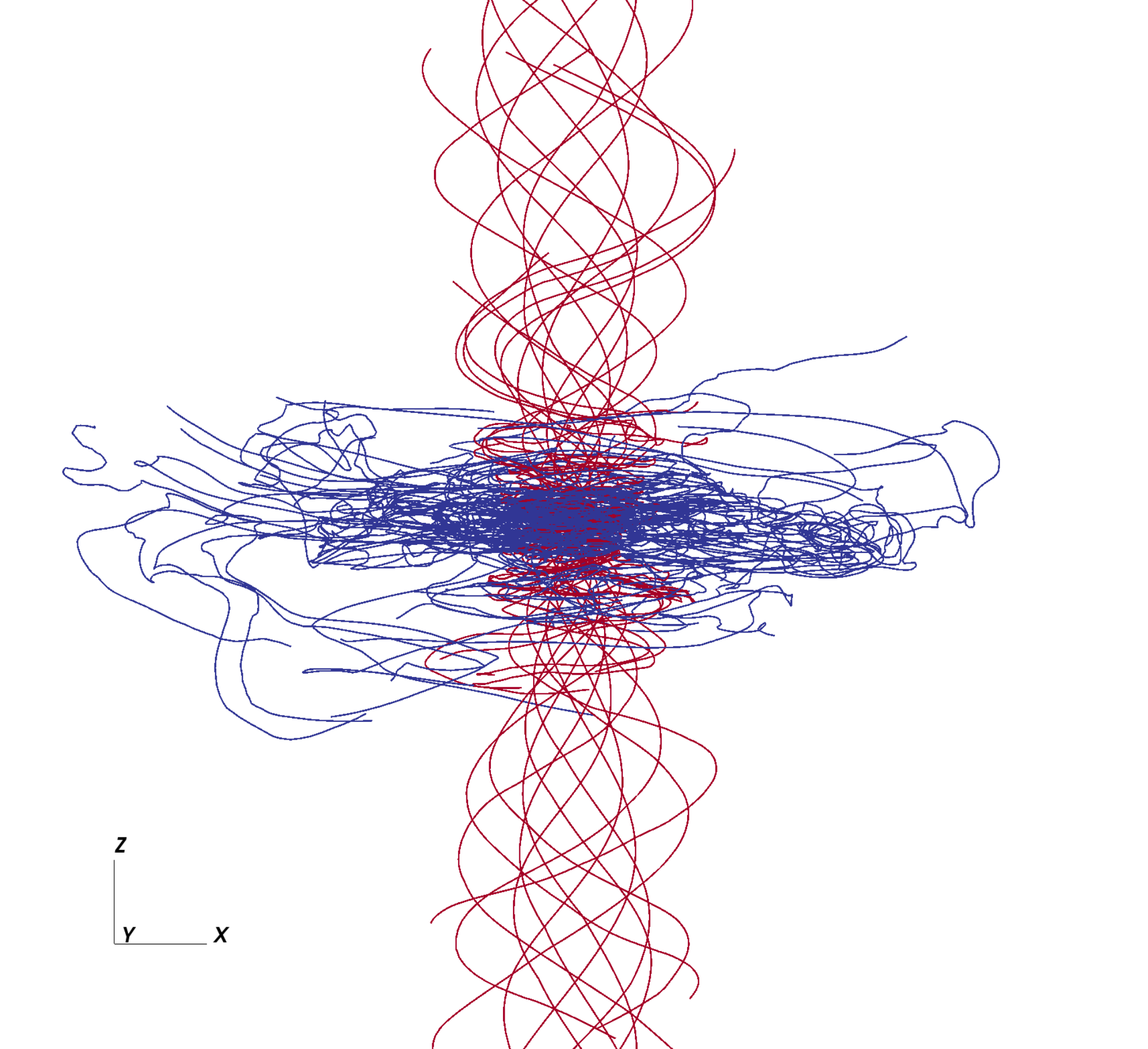}
  \end{tabular}
  \caption{Magnetic field line illustrations created with {\sc VisIt} by integrating magnetic field components in Cartesian Kerr-Schild coordinates.  The MAD $a_\bullet=+0.94$ simulation is visualised in the left column, while the SANE $a_\bullet=+0.94$ simulation is visualised in the right column.  Lines coloured red are anchored to a sphere of radius $2GM_\bullet/c^2$, while those coloured blue are anchored to a disk in the mid-plane.  The fields in the jet funnel are similar in both models, but the disk field is more ordered and has a stronger vertical component in the MAD model.  \label{fig:visit}}
\end{figure*}

By performing path integrals on the magnetic field (in Cartesian Kerr-Schild coordinates), we produce three-dimensional visualisations using {\sc VisIt}, shown in Figure \ref{fig:visit} \citep{VisIt2012}.  Magnetic field lines coloured blue are anchored within the plane of the disk, while magnetic field lines coloured red are anchored around the BH (on a sphere of radius $2 GM_\bullet/c^2$).  The left column depicts the MAD simulation, while the right column depicts the SANE simulation.  In both cases, the magnetic field in the funnel region is ordered and well-defined.  However, the disk magnetic field in the MAD model is more ordered and has a stronger vertical component than its SANE counterpart.  These visualisations also reveal a steep pitch angle in the funnel region, which affects the circular polarization produced by the transverse twist \citep[e.g.,][]{Gabuzda+2008}.  

\section{Circular Polarization Images}
\label{sec:circular_polarization}

\subsection{Circular Polarization in Our Models}
\label{sec:circular_polarization_intro}

\begin{figure*}
  \centering
  \includegraphics[width=\textwidth]{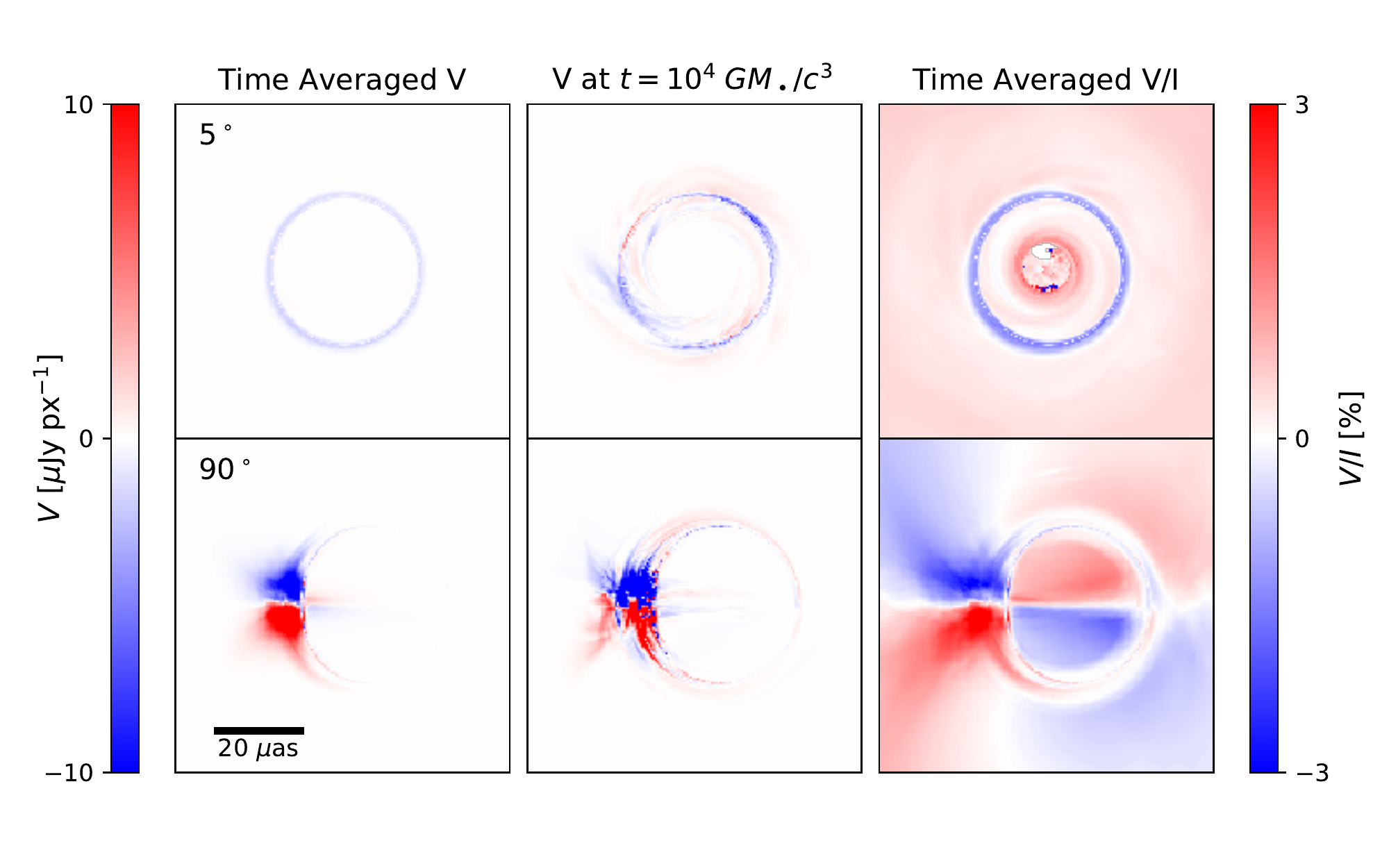}
  \caption{An introduction to the circular polarization of our MAD model.  The final snapshot ($t=10000 \ GM_\bullet/c^3$) is shown.  The top row displays the system with a pole-on inclination of $5^\circ$, while the bottom row displays the system at an edge-on inclination of $90^\circ$.  In the first and third columns, 501 images between $7500<GM_\bullet/c^3<10000$ have been averaged.  In the centre column, we display the final snapshot.  For this model, the photon ring dominates the circularly polarized image for pole-on inclinations, while a sign flip across the mid-plane that reflects a sign flip in the ``transverse twist'' is seen at edge-on inclinations.   \label{fig:intro_mad}}
\end{figure*}

\begin{figure*}
  \centering
  \includegraphics[width=\textwidth]{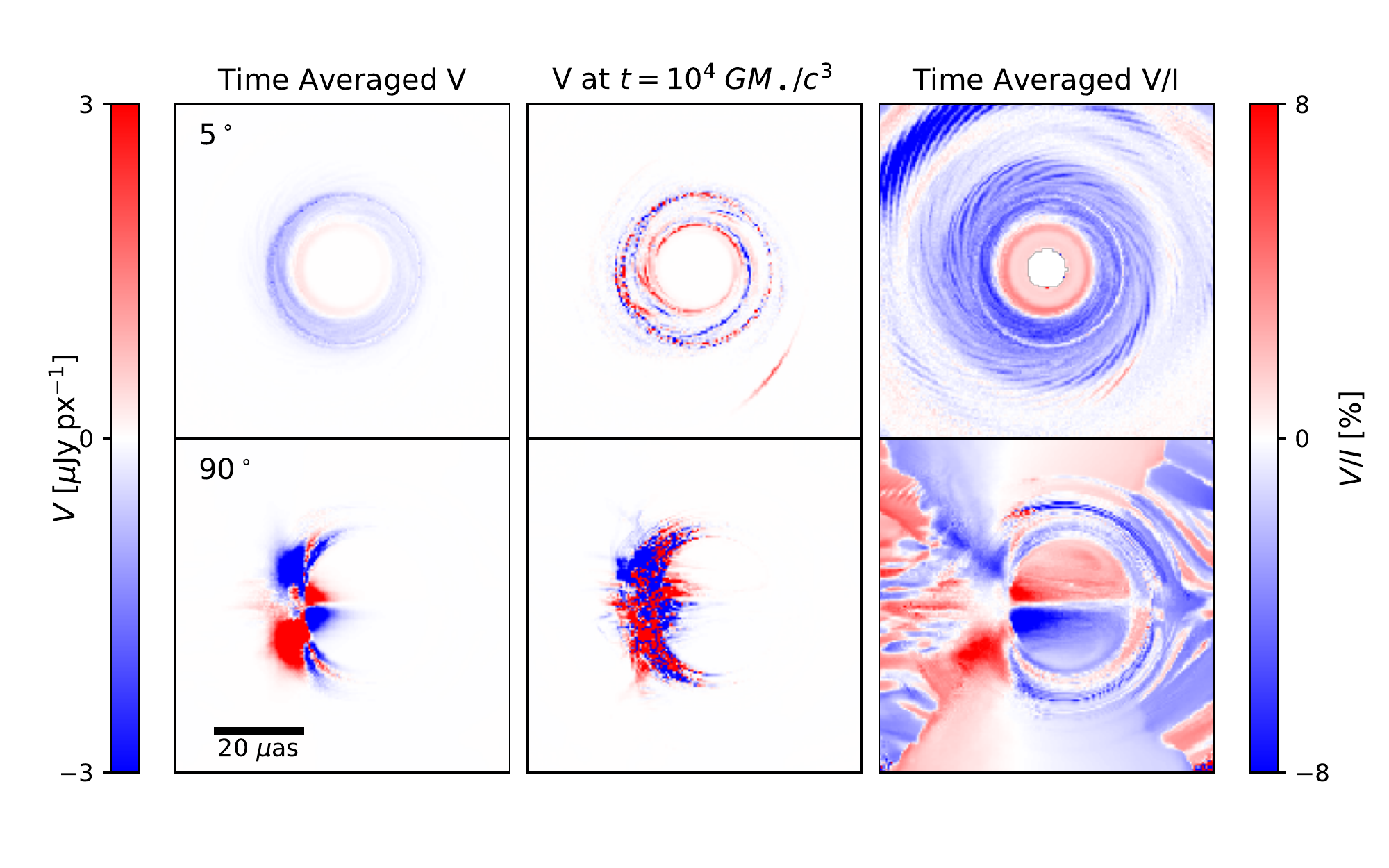}
  \caption{As Figure \ref{fig:intro_mad}, but for the SANE model.  At a given snapshot, the SANE model image is messier than the MAD's due to both its more turbulent disk and its higher Faraday depths.  Faraday rotation scrambles the photon ring more strongly in this model.  Note the different colour scalings for this figure compared to Figure \ref{fig:intro_mad}.  \label{fig:intro_sane}}
\end{figure*}

\begin{figure*}
  \centering
  \includegraphics[width=\textwidth]{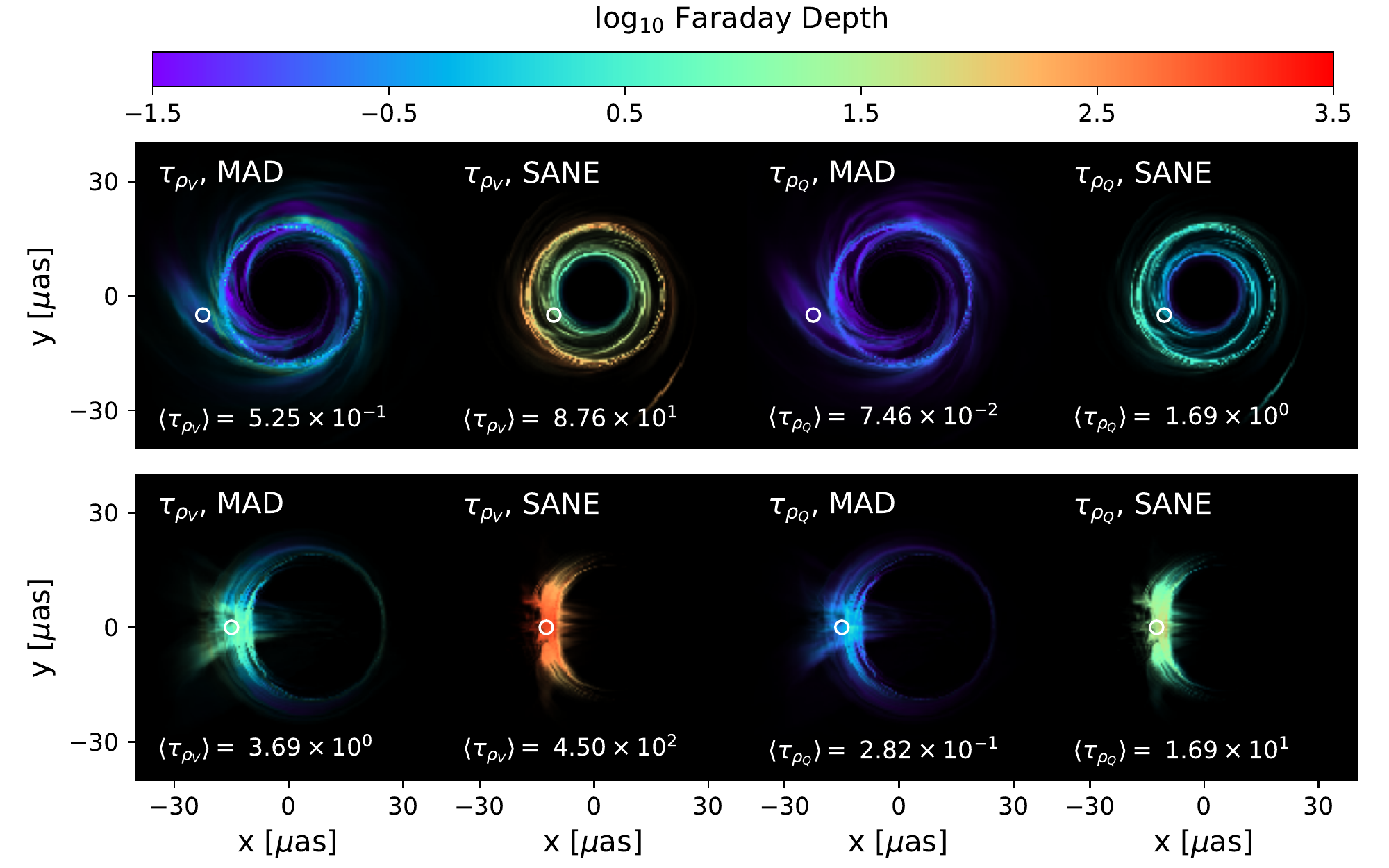}
  \caption{Depths of Faraday rotation ($\tau_{\rho_V}$) and conversion ($\tau_{\rho_Q}$) in our models when viewed at $5^\circ$ (top row) and $90^\circ$ (bottom row).  The MAD model is more Faraday thin than the SANE model.  Both models are more thick to Faraday rotation than Faraday conversion.  Models are more Faraday thick for edge-on inclinations than face-on ones. In the pixels marked with white circles, we study the emissivity and rotativity along the line of sight in Appendix \ref{sec:pixel_traces}.\label{fig:faraday_depths}}
\end{figure*}

In Figures \ref{fig:intro_mad} and \ref{fig:intro_sane}, we introduce the circular polarization images of the MAD and SANE models respectively.  The top row plots a pole-on viewing angle of $5^\circ$, while the bottom row plots an edge-on viewing angle of $90^\circ$.  For these figures alone, we have time-averaged 501 images, one for each of the snapshots in the final quarter of the simulation, $7500<GM_\bullet/c^3<10000$.  Time-averaged circular polarization images (Stokes $V$) are shown in the first column, the image of just the final snapshot is shown in the second column, and the circular polarization fraction (Stokes $V/I$) is shown in the third column.  Immediately we can see that images from both models and both viewing angles feature both positive and negative regions, which will complicate the interpretation of the unresolved Stokes $V$ measurements.  The face-on MAD image is surprisingly dominated by its photon ring (discussed at length in \S\ref{sec:magnetic_twist_faceon}), while the edge-on images are preferentially dominated by opposite signs of circular polarization on opposite sides of the mid-plane.

Comparing the two, the MAD image appears more ordered than its SANE counterpart.  Furthermore, the photon ring, prominent in the face-on MAD circular polarization image (and discussed at length in \S\ref{sec:magnetic_twist_faceon}), is suppressed in the SANE image.  Some of the disorder in the SANE is due to its more turbulent magnetic field structure, as shown in Figure \ref{fig:b_decomposition}.  Moreover, the SANE model is more scrambled by its much larger Faraday rotation and conversion depths.  {Further discussed in \S\ref{sec:inclination}, the much higher Faraday conversion depth in the SANE model also leads to a much larger fractional circular polarization \citep[see also][]{EHT8}.}  We plot the Faraday depths of these models in Figure \ref{fig:faraday_depths}, where the Faraday rotation depth $\tau_{\rho_V} = \int \rho_V ds$ and the Faraday conversion depth $\tau_{\rho_Q} = \int \rho_Q ds$.  Here, $\rho_V$ and $\rho_Q$ are the radiative transfer coefficients responsible for Faraday rotation and conversion respectively \citep[see][for more details]{Dexter2016,Moscibrodzka&Gammie2018}.  In this figure, the brightness of each pixel encodes its total intensity, while the colour encodes the appropriate Faraday depth as indicated in the colour bar.  The intensity-weighted image average of these values is written on the bottom of each panel.  The top row shows an inclination of $5^\circ$, while the bottom row shows an inclination of $90^\circ$.  In Appendix \ref{sec:pixel_traces}, we decompose the emissivity and rotativity along the lines of sight marked by white circles in this figure.

At $5^\circ$, the MAD model is both Faraday rotation and conversion thin, while the SANE model is strongly Faraday rotation thick and moderately Faraday conversion thick.  Faraday rotation is directly encoded in the rotation measure, which for M87* has been observed to have a magnitude of $\approx 10^5 \ \mathrm{rad} \; \mathrm{m}^{-2}$ and exhibit sign flips \citep{Goddi+2021}, consistent with GRMHD studies \citep{Ricarte+2020}.  At an inclination of $90^\circ$, the MAD model becomes moderately Faraday rotation thick, while the SANE becomes thick to both Faraday rotation and conversion.  We should thus expect that the circular polarization of the MAD model should more closely reflect the simple and ordered magnetic field structure discussed in \S\ref{sec:magnetic_field_decompositions}.  

Now, to understand the features of these images in detail, we will decompose their circular polarization images into components originating from intrinsic emission and Faraday conversion, and explore how the sign of the circular polarization encodes the magnetic field geometry.

\subsection{Intrinsic Emission}
\label{sec:intrinsic_emission}

To generate images of purely intrinsic emission, we remove Faraday conversion effects from {\sc IPOLE} by setting the radiative transfer coefficient $\rho_Q = 0$.  Without conversion, linear polarization cannot be transformed into circular, and the only remaining mechanism for the production of circular polarization is intrinsic synchotron emission.  Recall that the sign of the intrinsic emission encodes the direction of the line of sight magnetic field.

\subsubsection{Edge-on --- Four Quadrants}
\label{sec:intrinsic_emission_edgeon}

\begin{figure*}
  \centering
  \includegraphics[width=\textwidth]{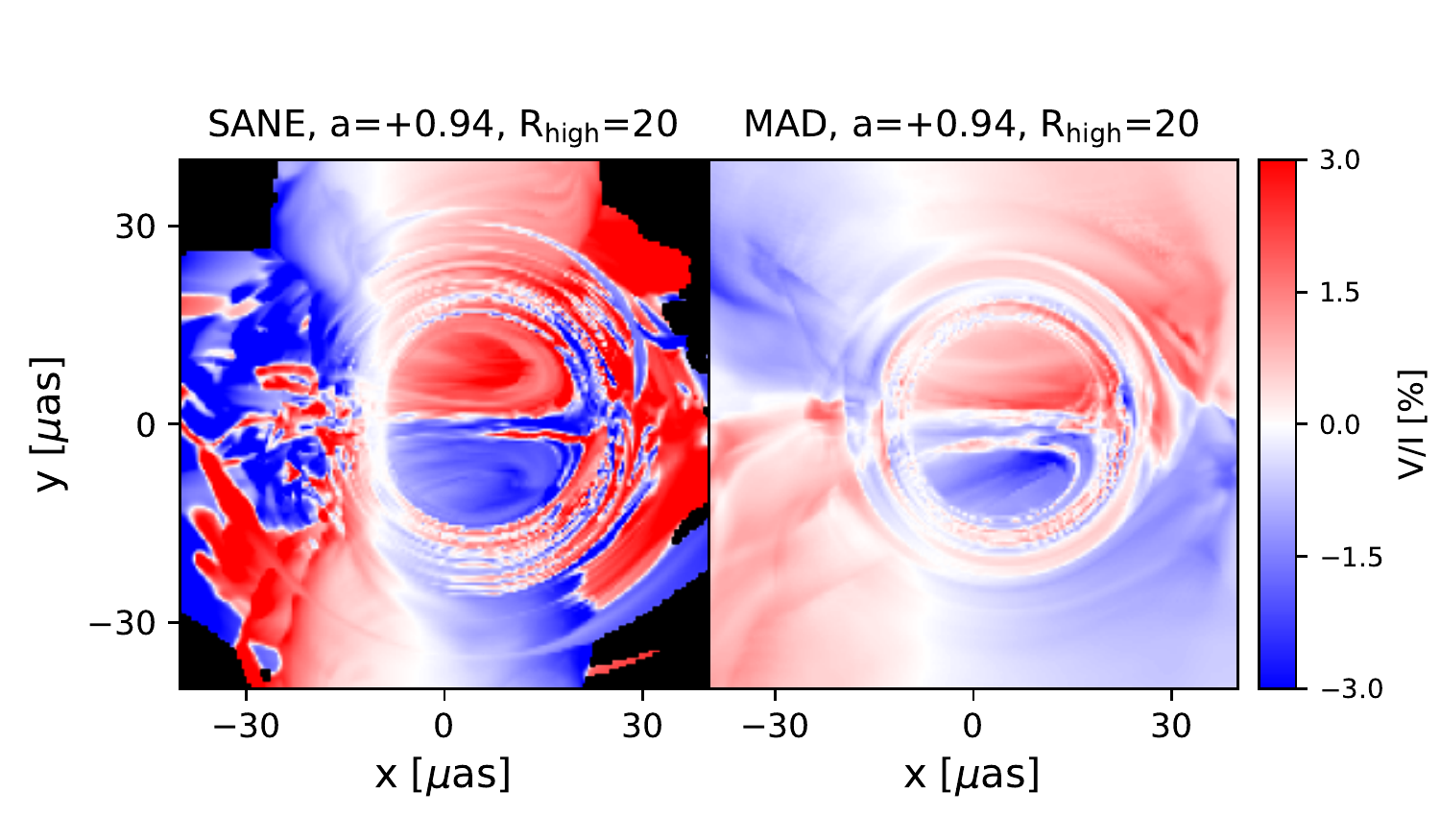}
  \caption{Intrinsic circularly polarized emission fractions (Faraday conversion switched off) for our two GRMHD models, viewed edge-on.  We see the ``four quadrants'' pattern resulting from the magnetic field geometry shown in Figure \ref{fig:cartoon_transverse_twist}.  Intrinsic emission dominates the circular polarization on large scales, producing a diffuse background. \label{fig:intrinsic_inc90}}
\end{figure*}

The geometry imprinted onto the intrinsic circular polarization is most clearly seen in the circular polarization fraction, $V/I$. This is plotted for our two models in Figure \ref{fig:intrinsic_inc90} for an edge-on viewing angle.  We see the ``four quadrants'' pattern of the magnetic field geometry as described in Figure \ref{fig:cartoon_transverse_twist}.  Recall that the sign flip across the $z$ axis is due to the helical twisting of field lines, while the sign flip across the mid-plane is physical.  The intrinsic circular polarization fraction can reach a few per cent in some areas.  However, circular polarization generated via intrinsic emission is sub-dominant in the bright parts of the image compared to that generated by Faraday conversion in these particular models, as we will explore in more detail in \S\ref{sec:inclination}.  In addition, we notice substantial cancellation due to the symmetry of these images.  The ``four quadrants'' pattern is also discussed in \citet{Tsunetoe+2020b}, who find that the leftward offset of the vertical sign-flip from the centre of the image is due to beaming effects.  \citet{Ricarte+2020} discuss the potential impact of bandwidth depolarization in images of Faraday thick models.  To ensure that bandwidth depolarization does not significantly impact our images, we average 129 individual images within a bandwidth of 4 GHz in Figures \ref{fig:intrinsic_inc90}, \ref{fig:intrinsic_inc5}, and \ref{fig:conversion_inc90}.  We found that bandwidth averaging has negligible effect on these images, owing in part to the relatively small value of $R_\mathrm{high}$ adopted.

\subsubsection{Face-on --- $B_z$, plus $B_r$ on the Far Side}
\label{sec:intrinsic_emission_faceon}

\begin{figure*}
  \centering
  \includegraphics[width=\textwidth]{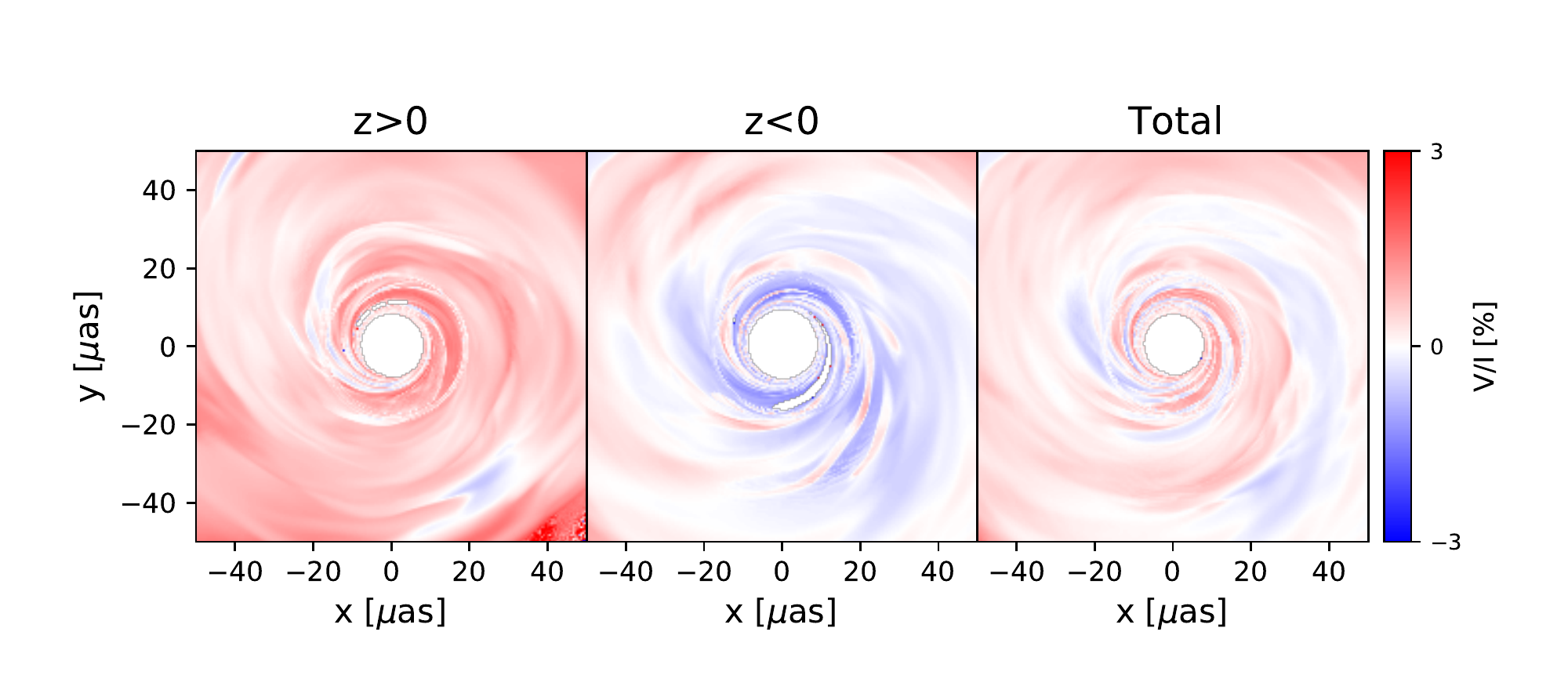}
  \caption{Intrinsic circularly polarized emission fractions (Faraday effects switched off) for the MAD model, viewed face-on.  One might expect a single sign of the circular polarization, corresponding to the vertical field direction.  Instead, we find a sign flip for emission originating on the far side of the mid-plane.  This is due to the inward bending of photon orbits, causing the circular polarization to be sensitive to the radial magnetic field in addition to the vertical component.  For the magnetic field geometry described in \S\ref{sec:magnetic_field_decompositions}, the radial field on the far side of the BH always has the opposite sign as the vertical field. \label{fig:intrinsic_inc5}}
\end{figure*}

For face-on inclinations, one might expect to see only a single sign of intrinsic circular polarization, corresponding to the direction of the vertical magnetic field, $B_z$.  Instead, as plotted in Figure \ref{fig:intrinsic_inc5}, we find that the intrinsic emission from the far side of the emitting region, behind the mid-plane, tends to exhibit the opposite sign.  This unanticipated phenomenon is due to the lensing of photon orbits.  Behind the mid-plane, photon orbits are bent towards the BH, causing the circular polarization to be sensitive to not only the vertical magnetic field, but also its radial component.  Assuming the geometry sketched in Figure \ref{fig:cartoon_transverse_twist}, $B_r$ on the far side always has the opposite sign of $B_z$.  Flips in the apparent line of sight direction of the magnetic field have also been found in these models in the context of Faraday rotation, which similarly encodes the direction of the magnetic field relative to the photon wavevector \citep{Ricarte+2020}.

\subsection{Faraday Conversion}
\label{sec:faraday_conversion}

Faraday conversion exchanges linear and circular polarization states.  On its own, it is insensitive to the sign of the magnetic field, and depends on the relative twist between the field producing the emission and the field performing the conversion.  We create images of circular polarization due only to Faraday conversion by switching off the coefficient responsible for intrinsic emission, $j_V=0$. In these particular models, Faraday conversion contributes more to the unresolved circular polarization image than the intrinsic emission, which we will further explore in \S\ref{sec:inclination}.

\subsubsection{Edge-on --- Transverse Twist}
\label{sec:magnetic_twist_edgeon}

\begin{figure*}
  \centering
  \includegraphics[width=\textwidth]{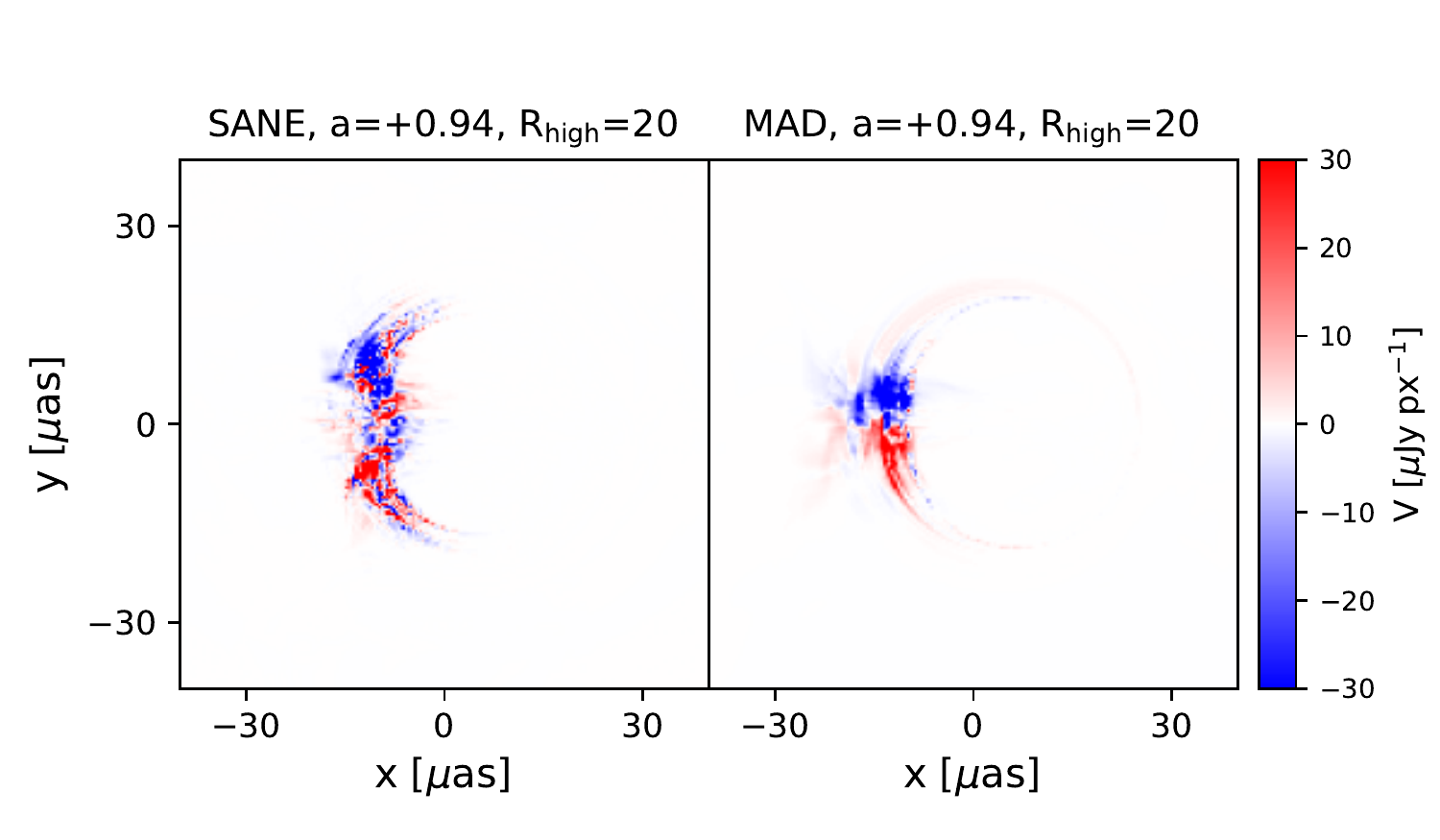}
  \caption{Circular polarization images of our two models, purely from Faraday conversion, $j_V=0$, from an edge-on viewing angle.  These images are very similar to the lower central panels of Figures \ref{fig:intro_mad} and \ref{fig:intro_sane}, because Faraday conversion dominates the circular polarization in the bright parts of the image.  Both images are predominantly negative on top and positive on the bottom, directly reflecting the direction of transverse twist shown in Figure \ref{fig:cartoon_transverse_twist}.
  \label{fig:conversion_inc90}}
\end{figure*}

For edge-on inclinations, the sign of the circular polarization encodes the transverse twist, $\xi_T$.  As illustrated in Figure \ref{fig:cartoon_transverse_twist}, $\xi_T$ flips sign at the mid-plane and depends on the direction that the magnetic field is twisted.  Circularly polarized intensity images created only by Faraday conversion are shown in Figure \ref{fig:conversion_inc90}.  These Stokes $V$ images are predominantly negative on top and positive on the bottom, as expected from the sign of $\xi_T$.  The SANE image is much messier due to both turbulence in the magnetic field and its thickness to both Faraday rotation and conversion.  Given a circularly polarized image of an edge-on system, we expect a sign flip at the mid-plane and for the signs to directly encode the handedness of the magnetic field on either side.

\subsubsection{Face-on --- Vertical Twist and a Photon Ring Signature}
\label{sec:magnetic_twist_faceon}

\begin{figure*}
  \centering
  \includegraphics[width=\textwidth]{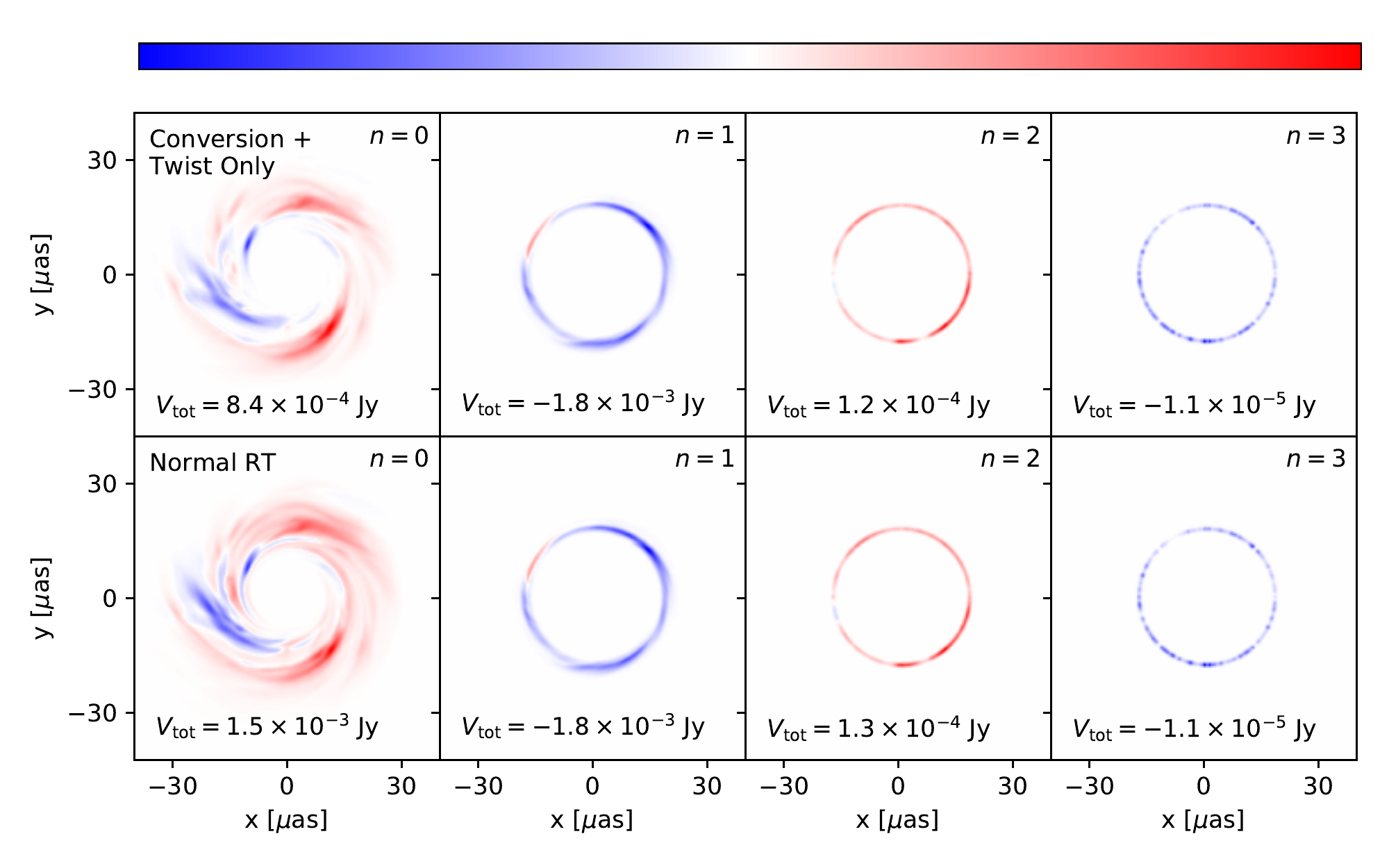}
  \caption{Circular polarization images of the MAD, $a_\bullet=0.94$, $R_\mathrm{high}=20$ model, broken into sub-ring components, revealing that alternating sub-rings flip sign.  We scale each panel's colour map individually such that the colours span $\pm \mathrm{max}|V|$.  On the top row, only circular polarization originating from Faraday conversion through the magnetic twist pathway is plotted, by setting the radiative transfer coefficients $\rho_V=j_V=0$.  All coefficients are active in the second row, demonstrating that Faraday conversion mediated by magnetic twist causes this signal.  Images are blurred with a Gaussian with a full-width at half-maximum of 1 $\mu$as to better visualise the narrow and faint sub rings.  Rings of increasing order alternate sign due to parallel transport and generic properties of the magnetic field geometry.  Under the right conditions, the circular polarization of the photon ring directly encodes the handedness (vertical twist) of the magnetic field.  All sub-images flip sign if the system is viewed from the opposite inclination, where the twists of the helices are reversed.
  \label{fig:photon_ring}}
\end{figure*}

For face-on viewing angles, circular polarization is sensitive to the vertical twist, $\xi_V$, which happens to have the opposite sign as the transverse twist.  In the MAD model, we notice the remarkable feature that the photon ring has the opposite sign compared to the rest of the image, a feature also discussed in \citet{Moscibrodzka+2021} for a different set of MAD simulations.  In fact, we find that photon sub-rings of increasing order alternate sign.  In Figure \ref{fig:photon_ring}, we break down the image of the final snapshot of this model into its photon ring components (George Wong, private communication).  To isolate the effect of Faraday conversion mediated by magnetic twist, Faraday rotation and intrinsic emission are turned off in the top row, but normal radiative transfer is performed in the bottom row.  In these panels, $n$ corresponds to the number of times the photon orbit has reached an extremum in the $\theta$ coordinate, counting the number of half-orbits the photon completes.  The total integrated Stokes $V$ for each sub-image is written on the bottom left.  Notice that the $n=0$ direct image has about a factor of two greater circular polarization when all coefficients are turned on compared to the pure conversion and twist image.  This behaviour matches expectations from a magnetic field pointed towards the observer, as we will further explore in \S\ref{sec:inclination}.   For the photon ring and its sub-rings, $n\geq 1$, Faraday conversion mediated by magnetic twist completely explains the signal.  Remarkably, the $n=1$ photon ring dominates the circularly polarized image in this model, even though it contributes to only 16 per cent of the total (unpolarized) flux, and we confirm this to be generally true for snapshots in times $7500<GM_\bullet/c^3<10000$.  As has been shown in previous works, photon rings of increasing order have intensities which decrease by about an order of magnitude with each half-orbit \citep{Johnson+2020,Himwich+2020}, but the flips in the sign of Stokes V are unanticipated.

\begin{figure*}
  \centering
  \includegraphics[width=\textwidth]{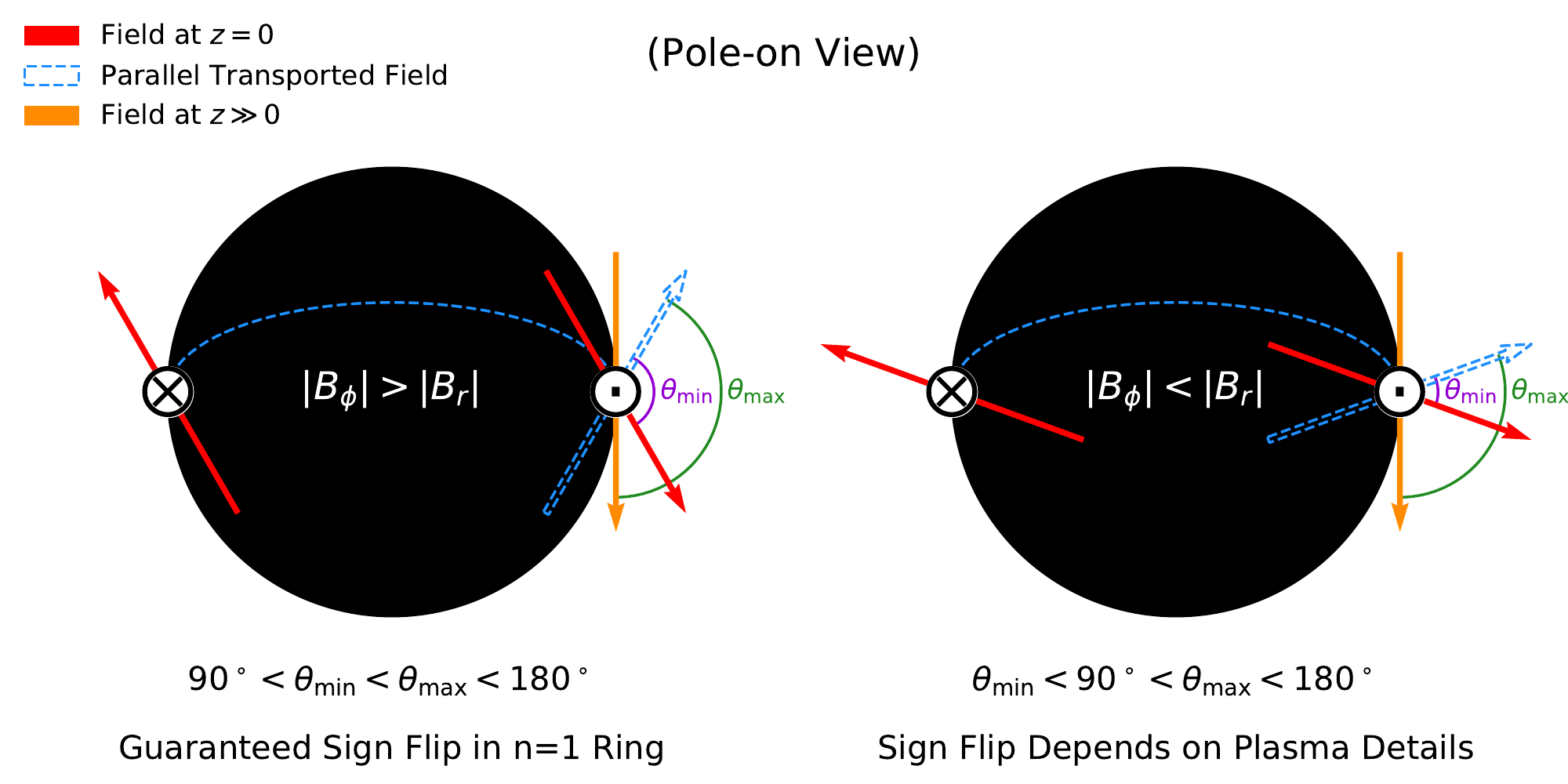}
  \caption{Schematic explaining the sign flips in the photon ring for a face-on viewing angle due to magnetic field symmetries and parallel transport.  Here, we consider a pixel on the right side of the photon ring (where the arrows intersect).  The magnetic field in the mid-plane is shown in red, while the field at $z \gg 0$ (towards the camera) is shown in orange.  (These are the magnetic fields depicted in Figure \ref{fig:cartoon_vertical_twist}.)  Emission in the $n=1$ photon ring is loaded on the opposite side of the BH, so to determine the line of sight twist, we parallel transport the emitting field along a photon geodesic around the SMBH, as illustrated in blue.  The sign of Stokes $V$ depends on the angle that this emitting field makes with the converting field ahead of it, and this is bracketed by $\theta_\mathrm{min}$ and $\theta_\mathrm{max}$.  For a toroidally dominated field (left), these angles are always confined to the quadrant between $90^\circ$ and $180^\circ$, which guarantees a sign flip in the $n=1$ ring.  This is not the case for a radially dominated field (right), so the existence of a sign flip depends on exactly where along the line of sight the Faraday conversion occurs.  Faraday rotation can also disrupt these inequalities and corrupt this signal.
  \label{fig:parallel_transport_schematic}}
\end{figure*}

These sign flips can be explained by a combination of symmetries of the magnetic field, the dominance of its azimuthal component, and features of parallel transport.  We illustrate how this results in a sign flip in the photon ring in Figure \ref{fig:parallel_transport_schematic}.  We consider a pole-on viewing angle and a pixel which appears in the right side of the photon ring (at the intersection of the arrows).  By construction, intensity in the $n=1$ photon ring component originates from the opposite side of the SMBH, having bent around the back side (behind the page).  The field in the mid-plane is represented by red arrows, while the field at $z \gg 0$ (towards the observer) is represented by the orange arrow.  The field at $z \gg 0$ is twisted following the rules described in Figure \ref{fig:cartoon_vertical_twist}.  In the left panel, the magnetic field is toroidally dominated (as in our models), while in the right panel, the field is radially dominated.

To determine the magnetic twist along this bent trajectory, we must parallel transport the emitting field (on the left, moving in the $\bigotimes$ direction) along the geodesic to the converting field (on the right, moving in the $\bigodot$ direction).  When parallel transported around the back of the SMBH, the azimuthal component of the magnetic field flips sign, but the radial component does not.\footnote{For spinning SMBH, a photon will also move in the azimuthal direction, but due to the axissymmetry of the problem, this has no effect on our argument.}  The relative twist between photons loaded on the left side of the SMBH and the Faraday converting material on the right side can be inferred by comparing the parallel transported field line (as shown in blue) with the field orientations both in the mid-plane (red) and at larger distances towards the camera (orange).  We can see that this brackets the twist angles $\theta_\mathrm{min}$ (at the mid-plane) and $\theta_\mathrm{max}$ (at infinity).  

Next, recall that Stokes $V$ inherits the sign of Stokes $U$, and Stokes $U$ depends on the relative twist between the emitting field and the converting field.  In the plasma frame, synchrotron emission produces light polarized in the $+Q$ orientation, but a twist in the magnetic field in the propagation direction can recast what was once entirely $+Q$ into some parts $\pm U$.  Most importantly, the resultant sign of $U$ flips each time the twist increases by $90^\circ$.  For the toroidally dominated field on the left, the twist angle is always between 90$^\circ$ and 180$^\circ$.  Therefore, the $n=1$ photon ring should always exhibit the opposite sign of Stokes $V$ compared to the $n=0$ image overall.  Generalizing to $n>1$, emission in odd numbered sub-rings is generated on the left side of this schematic, while emission in even numbered sub-rings is generated on the right side.  Consequently, odd numbered and even numbered sub-rings should each share signs.

If the field is radially dominated, as shown on the right, then unfortunately the sign of the photon ring depends on the details of the plasma.  Parallel transported around the SMBH, the magnetic field would intersect the magnetic field near the mid-plane at an angle $\theta_\mathrm{min}<90^\circ$, but would eventually encounter magnetic field where the relative angle $\theta_\mathrm{min}>90^\circ$.  The resultant sign of the circular polarization then depends on where exactly the Faraday conversion occurs along this trajectory.

To summarise, for face-on viewing angles, the sign of circular polarization of the photon ring can directly encode the handedness (the vertical twist) of the magnetic field in the approaching helix.  If viewed from the opposite inclination, all sub-images would flip sign, since the handedness of the field flips across the mid-plane.  As long as Faraday rotation is weak, sign flips in Stokes $V$ are guaranteed in sub-images of increasing sign for toroidally dominated fields, and may also occur in radially dominated fields depending on the plasma details.  This signal can be explained by generic properties of the magnetic field geometry, the sensitivity of circular polarization to both the direction and degree of magnetic twist, and features of parallel transport within the photon ring.  EHT imaging constraints favour clockwise rotation of material in the accretion flow of M87* \citep{EHT5}.  If the accretion flow is produced by a prograde model with $|B_\phi| > |B_r|$, then the photon ring should have $V>0$.  If the accretion flow is retrograde, we anticipate additional complications, as we discuss in \S\ref{sec:retrogrades}.  

If Faraday rotation is strong, it can corrupt this signal in two ways.  First, Faraday rotation along the line of sight can further rotate the plane of polarization based on the line of sight magnetic field direction, potentially disrupting the inequalities in Figure \ref{fig:parallel_transport_schematic}.  Indeed, we caution that for our Faraday rotation thick SANE model, we only see the signal exactly as described in this section if Faraday rotation is switched off.  Second, strong enough Faraday rotation can scramble the circular polarization by randomising the sign of Stokes $U$ that gets converted to $V$, which especially affects photon ring geodesics, since they have larger than usual path lengths through the system.  Complementary linearly polarized images, especially at multiple frequencies, would help assess if Faraday rotation in the photon ring is important.

\subsection{Unresolved Circular Polarization Measurements}
\label{sec:inclination}

\begin{figure*}
  \centering
  \includegraphics[width=\textwidth]{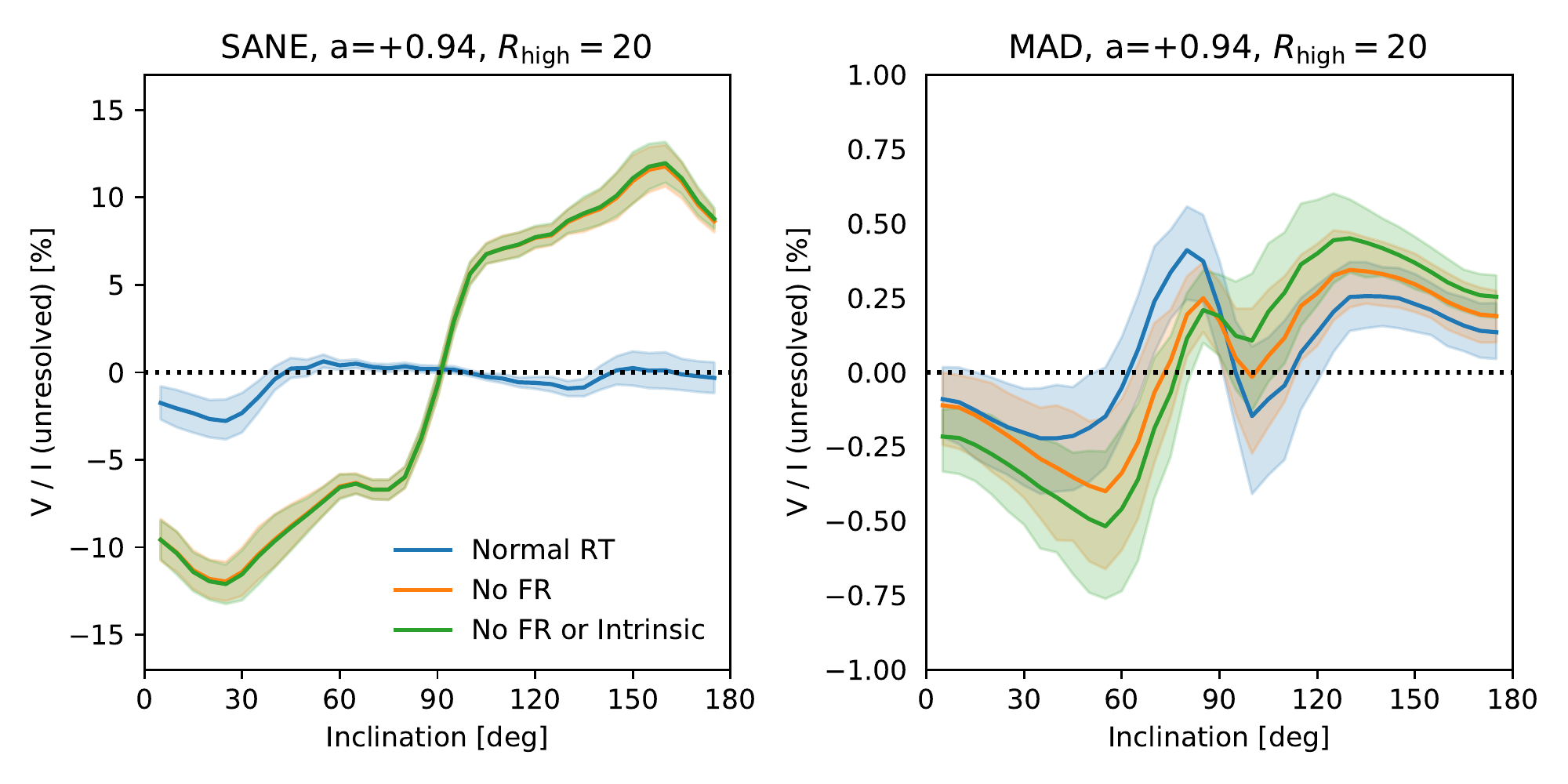}
  \caption{Image-integrated circular polarization fractions as a function of inclination for our two models.  Normal radiative transfer is plotted in blue, images created without Faraday rotation are plotted in orange, and images created without Faraday rotation or intrinsic emission are plotting in green.  We include multiple snapshots in this plot, and the filled regions enclose the 5th to 95th percentiles of the values spanned between times $t \in [7500,10000] \ GM_\bullet/c^3$, while the solid lines mark the median.  This reveals that constancy of the sign of $V/I$ occurs at some, but not all, inclinations.  These models exist in two different regimes of Faraday rotation.  In the SANE model, Faraday rotation is strong enough to significantly depolarize the circularly polarized image, while in the MAD model Faraday rotation is weak enough to imprint the magnetic field direction on the sign of $V/I$, pushing it towards positive values for inclinations $<90^\circ$ and towards negative values for inclinations $>90^\circ$.  Intrinsic emission imprints the line of sight direction of the magnetic field in a similar fashion. \label{fig:unresolved_V}}
\end{figure*}

The sign of unresolved circular polarization measurements has been proposed to infer the sense of rotation in quasar engines \citep{Ensslin+2003}.  Using our two models, we probe to what extent the unresolved circular polarization fraction and its time variability can be used to infer the magnetic field geometry on event horizon scales.  In Figure \ref{fig:unresolved_V}, we compute unresolved circular polarization fractions for each of our models stepping through inclination from $5^\circ$ to $175^\circ$ in $5^\circ$ increments.  In addition to sampling different inclinations, we also make separate images for the times $7500 \ GM_\bullet/c^3$ to $10000 \ GM_\bullet/c^3$, sampled in steps of $100 \ GM_\bullet/c^3$.  For a given model at a given inclination, the filled region encloses the 5th to 95th percentiles of the range spanned by these models as a function of time, while the solid line marks the median.  This reveals that constancy of the sign of $V/I$ over time occurs at some, but not all, inclinations for each model.  In blue, we plot the results of normal radiative transfer (Normal RT), in orange we switch off Faraday rotation (No FR), and in green we switch off both Faraday rotation and intrinsic emission of circular polarization (No FR or Intrinsic).  

The surprising amount of structure in the circular polarization curves as a function of inclination reflects the interplay of the three pathways to generate circular polarization.  Even when facing the same side of the disk, it is possible to obtain both positive and negative circular polarization fractions.  The ``Normal RT'' curve is not perfectly anti-symmetric about $90^\circ$ because not all physical effects depend on inclination in the same way.  

By comparing the blue ``Normal RT'' curves with the orange ``No FR'' curves, we see two regimes in which Faraday rotation affects the circularly polarized image.  In the MAD model, which has mild values of $\tau_{\rho_V}$, Faraday rotation shifts circular polarization fractions towards positive values for inclinations $<90^\circ$, and towards negative values for inclinations $>90^\circ$.  This is because Faraday rotation is sensitive to the sign of the magnetic field, which is oriented vertically upwards in these models.  In contrast, the SANE model is much more Faraday rotation thick, and hence Faraday rotation strongly suppresses the circular polarization, which in its absence could grow to values as large as 10 per cent.  This is because, as shown in Figure \ref{fig:faraday_depths}, the SANE model is much more Faraday conversion thick than the MAD model. Such a large circular polarization fraction is strongly ruled out by current constraints on M87* \citep[$\lesssim 0.8$ per cent][]{Kuo+2014,Goddi+2021}.  For some inclinations, this internal Faraday rotation can even change the sign of the unresolved circular polarization fraction.  Finally, intrinsically emitted circular polarization modifies these curves in a similar direction by adding emission whose sign reflects the line of sight magnetic field direction, just like a small amount of Faraday rotation.

In summary, we find that the sign of the unresolved circular polarization fraction cannot easily be used to infer knowledge of the system without additional constraints.  All three pathways to generate circular polarization are relevant and depend on inclination in different ways.  Both models can exhibit constancy of the sign of the $V/I$ with time at some, but not all, inclinations.

\section{Discussion}
\label{sec:discussion}

\subsection{Dipolar vs. Quadrupolar Fields}
\label{sec:quadrupolar_fields}

The models considered in this work have dipolar magnetic fields, while a quadrupolar magnetic field is also possible \citep[e.g.,][]{Beckwith+2008}.  A spun-up quadrupolar magnetic field would not exhibit a sign-flip in the tangential and radial components of the magnetic fields at the mid-plane.  While the twist of these magnetic field structures would be qualitatively similar to that of a dipolar field, their field directions would not be.  Hence, if the field were quadrupolar rather than dipolar, the intrinsic emission and Faraday rotation would be affected, but not the emission due to Faraday conversion.  Since Faraday conversion dominates our images, we expect that circularly polarized images created with models with quadrupolar fields should be qualitatively similar to those created with dipolar fields.

\subsection{Electron-Positron Plasmas}
\label{sec:pair_plasmas}

A substantial population of positrons may exist within these accretion flows via pair production \citep{Begelman+1984,Reynolds+1996b,Wardle+1998}, which would modify the equations of polarized radiative transfer.  As the ratio of positrons to electrons approaches unity, intrinsically emitted circular polarization and Faraday rotation vanish, while Faraday conversion doubles \citep{Park&Blackman2010,Ensslin+2019,Anantua+2020,Emami+2021}.  Consequently, the existence of a positron population would result in images which would more directly encode the twist in the magnetic field geometry.  In particular, the existence of positrons could strengthen the signature of magnetic twist in the photon ring described in \S\ref{sec:magnetic_twist_faceon}.

\subsection{Retrograde Models}
\label{sec:retrogrades}

\begin{figure*}
  \centering
  \includegraphics[width=\textwidth]{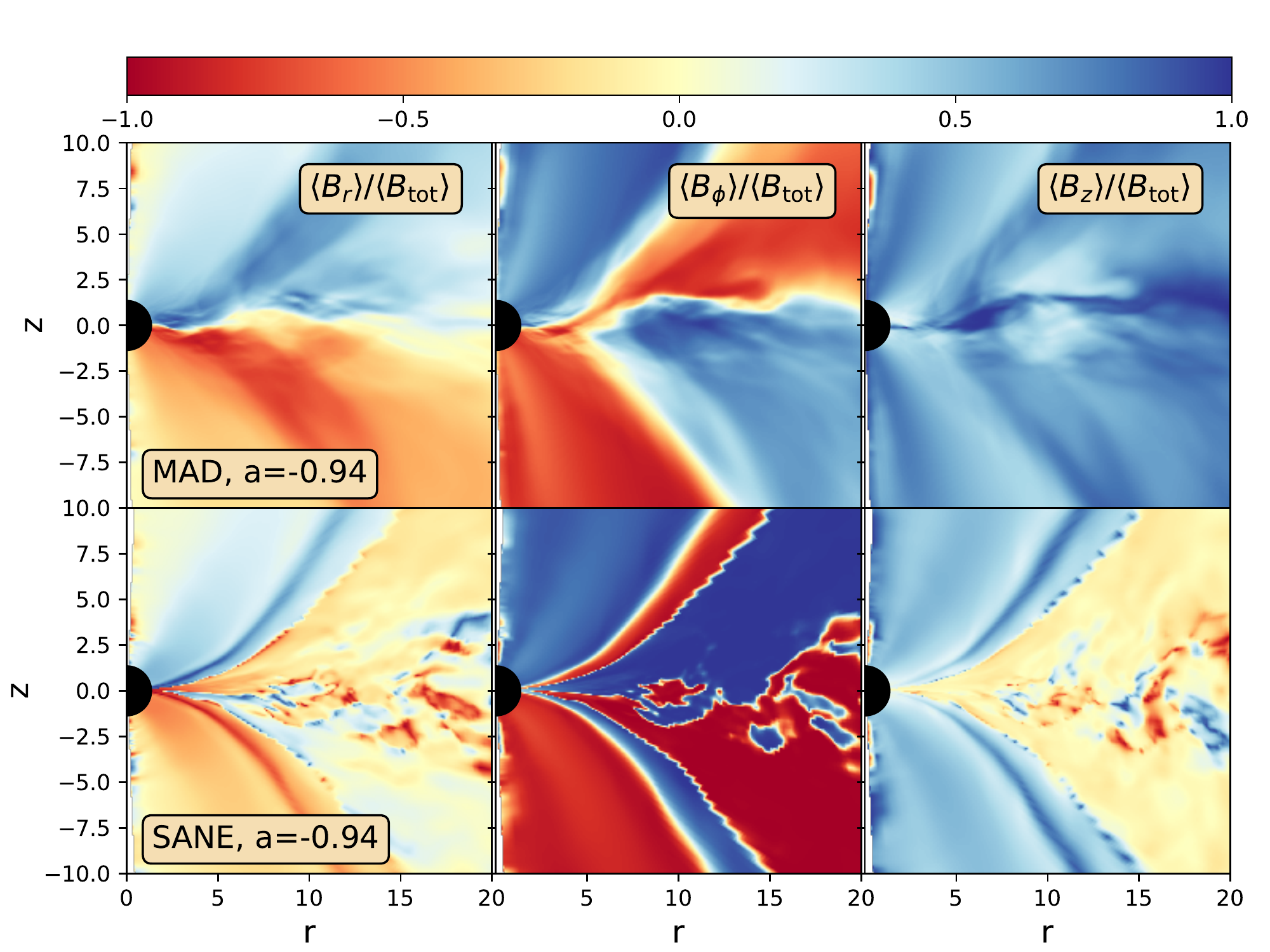}
  \caption{Magnetic field decompositions as in Figure \ref{fig:b_decomposition}, but for retrograde models.  In the funnel region, the magnetic field aligns with the SMBH.  Beyond this region, in the MAD model, the field neatly switches azimuthal direction to align with the disk instead.  However, in the SANE model, the field lines only align with the disk angular momentum in a narrow shell around the funnel.  These additional features may produce more complicated circular polarization images depending on the details of how they are illuminated. \label{fig:b_decomposition_retrogrades}}
\end{figure*}

In a retrograde model, frame dragging operates in the opposite direction as the disk angular momentum, which produces more complicated magnetic field geometries than sketched in Figure \ref{fig:cartoon_vertical_twist}.  In Figure \ref{fig:b_decomposition_retrogrades}, we plot the magnetic field decompositions of two retrograde models as in Figure \ref{fig:b_decomposition}.  In this plot, the $z$-axis remains aligned with both the vertical magnetic field direction and the disk angular momentum, but is now anti-aligned with the angular momentum of the SMBH.  The magnetic field in the funnel region is twisted along with the SMBH angular momentum.  In the MAD model, the field outside this region is twisted along with the disk, producing opposite twist orientations and directions of the toroidal field.  Meanwhile, in the SANE model, we find that the azimuthal field direction only consistently follows the disk angular momentum in a thin shell around the funnel, beyond which the field becomes more turbulent, and in this model in alignment with the SMBH angular momentum again.  We verify that this unexpected behaviour persists throughout times $t \in [7500,10000] \ GM_\bullet/c^3$.  As a result, we expect that the imprint of this structure on circular polarization images would likely be more sensitive to the details of the plasma properties---where these transitions in magnetic field orientation occur, where the emission occurs, and where the Faraday effects occur.  

Interestingly, the asymmetry of the image of M87* combined with its known jet orientation has constrained the motion of plasma on the sky as clockwise, and the orientation of the SMBH spin direction as away from us \citep[see][for an extended discussion]{EHT5}.  If we can assume that the evacuated inner funnel region does not contribute much to the radiative transfer, then it is the angular momentum of the disk in the boundary region that matters when determining the twist, rather than that of the SMBH.  Consequently, if M87*'s disk is prograde, we expect its photon ring to exhibit positive $V$, while if the disk is retrograde, we expect negative $V$ instead.  Future studies thoroughly exploring a wider variety of GRMHD and radiative transfer models would be useful in determining whether or not the sign of the circular polarization of the photon ring is a clean signature of prograde versus retrograde accretion.

\subsection{Lower Frequencies and Larger Spatial Scales}
\label{sec:other_frequencies}

Farther from the horizon, it is more straightforward to infer magnetic field properties free from the complications of event horizon scale magnetic field structure and light bending.  If measurable, the combination of a rotation measure gradient across the jet, the pitch angle inferred from linear polarization ticks, and the sign of circular polarization can be used to determine the directions of both the toroidal and poloidal field components \citep{Gabuzda+2008,Gabuzda2018a}.  Linear polarization maps exist for M87* at 43 GHz with sub-milliarcsecond resolution, and a full rotation measure analysis of this dataset is forthcoming \citep{Walker+2018}.  This information could help us associate regions of a resolved circularly polarized image in the millimetre to the photon ring and the different sides of the mid-plane.

\subsection{Non-thermal Electron Distributions}
\label{sec:nonthermal}

This study was performed using only a single model of the electron distribution function \citep{Moscibrodzka+2016}, distributed thermally.  Meanwhile, the SEDs of these sources imply the existence of non-thermal electron populations to produce flux particularly in the near infrared \citep{Ozel+2000,Yuan+2003}.  Images including a non-thermal component to the electron distribution function in the form of a high-energy tail exhibit wider halos of emission than purely thermal models \citep{Ozel+2000,Mao+2017}.  Since we find that images of circular polarization on large scales are dominated by intrinsic emission, we hypothesise that the intrinsic emission signatures discussed in \S\ref{sec:intrinsic_emission} may be strengthened.  For face-on inclinations, this would preferentially add circular polarization with the line of sight sign of the magnetic field, while for edge-on inclinations, this would strengthen the ``four quadrants'' pattern.  We expect that additional Faraday effects would be weaker than the additional intrinsic emission, since Faraday rotation and conversion are inefficient for high-energy electrons.  Light bending complications should also be weaker for a diffuse halo component.  An additional non-thermal component at lower energies would increase Faraday effects, however.  Overall, the effect of non-thermal electrons on circular polarization merits further study, and tractable prescriptions have been developed for the electron distribution function as a function of plasma $\beta$ and magnetization $\sigma$ that can be explored \citep{Ball+2018,Davelaar+2018,Davelaar+2019}.  

\section{Conclusions}
\label{sec:conclusion}

Using two GRMHD simulations of prograde accretion flows, we have studied the magnetic field structure and its imprint on circular polarization images.  Our results are summarised as follows:

\begin{itemize}
    \item First, we explore the line of sight direction and twist of the magnetic field in these models, which become imprinted on the circular polarization signal.  The handedness of the magnetic twist flips across the mid-plane.  In addition, we find that sign of the magnetic twist is different depending on whether the helix is viewed at edge-on versus pole-on viewing angles.
    \item Intrinsically emitted circular polarization carries the sign of the line of sight direction of the magnetic field.  For edge-on viewing angles, this results in a ``four quadrants'' pattern, also reported by \citet{Tsunetoe+2020b}.  For face-on viewing angles, this mainly traces the vertical magnetic field direction, but is also sensitive to the radial component of the magnetic field behind the BH due to the lensing of photon orbits.  Intrinsic emission produces relatively diffuse emission on large scales in these models.
    \item Circular polarization from Faraday conversion inherits the sign of Stokes $U$, which we find is sensitive to the line of sight twist in the magnetic field.  For edge-on viewing angles, we expect a sign flip across the mid-plane due to the switch in handedness of the magnetic twist.  For pole-on viewing angles, we find that alternating sub-images in the photon ring can flip sign.  This is due to parallel transport and symmetries of the magnetic field.
    \item As discussed in previous works, a mild amount of Faraday rotation will imprint the line of sight direction of the magnetic field on the circular polarization \citep{Tsunetoe+2020a,Tsunetoe+2020b}.  However, too large an amount of Faraday rotation will scramble the circular polarization, even if the model is ``thin'' to Faraday conversion.  We notice the former in our MAD model, and the latter in our SANE model.
    \item The circular polarization fraction and sign changes non-trivially with the inclination due to the complicated magnetic field geometry, different competing physical processes, and light bending.  For a given model, it is possible to obtain either sign of circular polarization while viewing the same side of the disk.
\end{itemize}

This work represents an early step in decomposing and understanding circularly polarized images on event horizon scales.  In future work, it will be useful to explore to what extent our findings generalise using different GRMHD and radiative transfer models.  In particular, we plan on expanding our analysis to study the impact of non-thermal electron distributions and positron populations.

\section{Acknowledgements}

We thank Charles Gammie, George Wong, and Ben S. Prather for support using their GRMHD simulations, {\sc ipole}, and {\sc VisIt}. We also thank Michael Johnson, Daniel Palumbo, Elizabeth Himwich, and Zachary Gelles for many fruitful and invigorating discussions about polarimetry, general relativity, and magnetic fields.  

This material is based upon work supported by the National Science Foundation under Grant No. OISE 1743747.  Computations at the Black Hole Initiative (BHI) were enabled by grants from the Gordon and Betty Moore Foundation and the John Templeton Foundation. 

\section{Data Availability}

The data underlying this article will be shared on reasonable request to the corresponding author.

\bibliography{ms}

\appendix

\section{Confirming Vertical Twist in our GRMHD Models}
\label{sec:app_vertical_twist}

\begin{figure}
  \centering
  \includegraphics[width=0.5\textwidth]{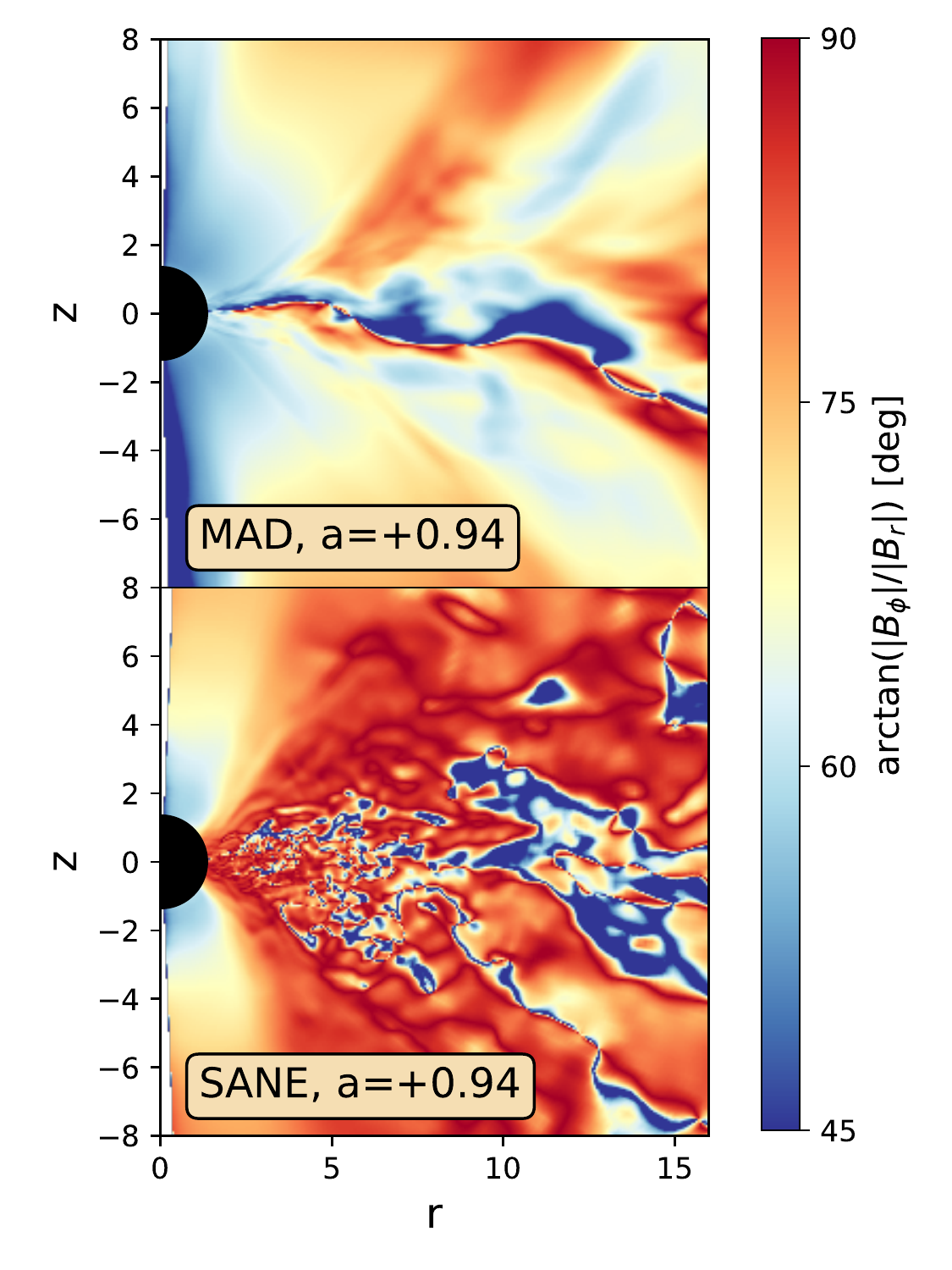}
  \caption{By examining the angle of the magnetic field projected onto the $xy$-plane, we can affirm the sense of vertical twist illustrated in Figure \ref{fig:cartoon_vertical_twist}.  Here, an angle of 45$^\circ$ implies that $|B_\phi| = |B_r|$, while an angle of 90$^\circ$ implies that $|B_\phi| \gg |B_r|$.  At least in the funnel region, this angle increases as $|z|$ increases.  This is not true in the disk of the turbulent SANE, where indeed our time-averaged $V$ image reveals weak edge-on polarization.  \label{fig:b_toroidal_angle}}
\end{figure}

In Figure \ref{fig:b_toroidal_angle}, we confirm the sense of ``vertical twist'' sketched in Figure \ref{fig:cartoon_vertical_twist}.  Ignoring the vertical component, we plot the angle made by the magnetic field in the $xy$ plane, $\arctan{|B_\phi|/|B_r|}$.  An angle of $45^\circ$ implies that the radial and azimuthal fields are equal in magnitude, while an angle of $90^\circ$ implies that the toroidal field is much stronger.  At fixed $r$, this angle increases as one moves away from the mid-plane, resulting in the vertical twist shown in Figure \ref{fig:cartoon_vertical_twist}.  However, this is only clear in the funnel region of the SANE model, outside of which the field remains consistently toroidally dominated and turbulent.  The vertical twist can also be inferred by close inspection of the pole-on viewing angles in Figure \ref{fig:visit}.  The red curves that are anchored around the BH, which tend to extend to larger $|z|$, sweep more circular arcs than the blue curves anchored in the disk. 

\section{Emissivity and Rotativity in Representative Pixels}
\label{sec:pixel_traces}

\begin{figure}
  \centering
  \includegraphics[width=0.5\textwidth]{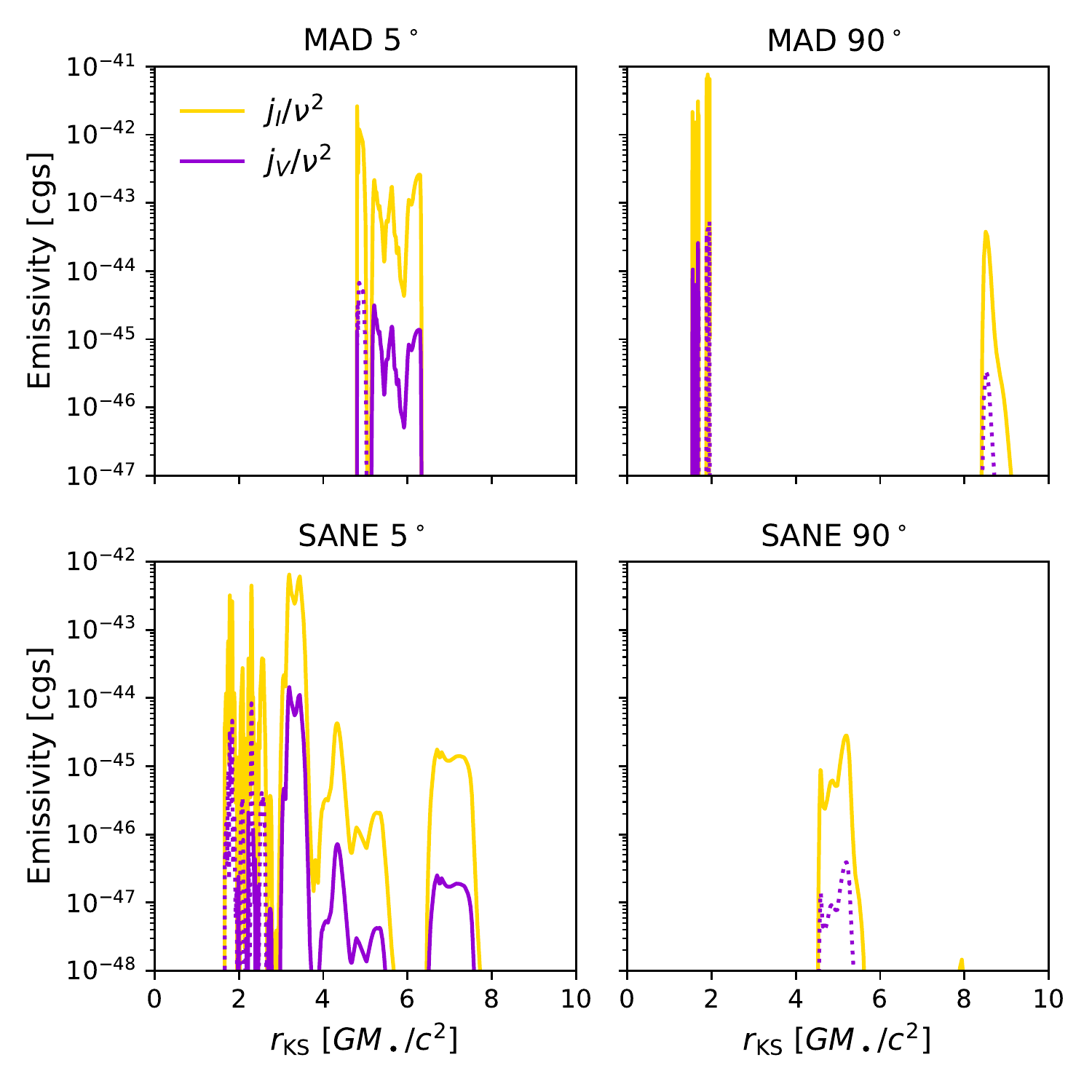} \\
  \includegraphics[width=0.5\textwidth]{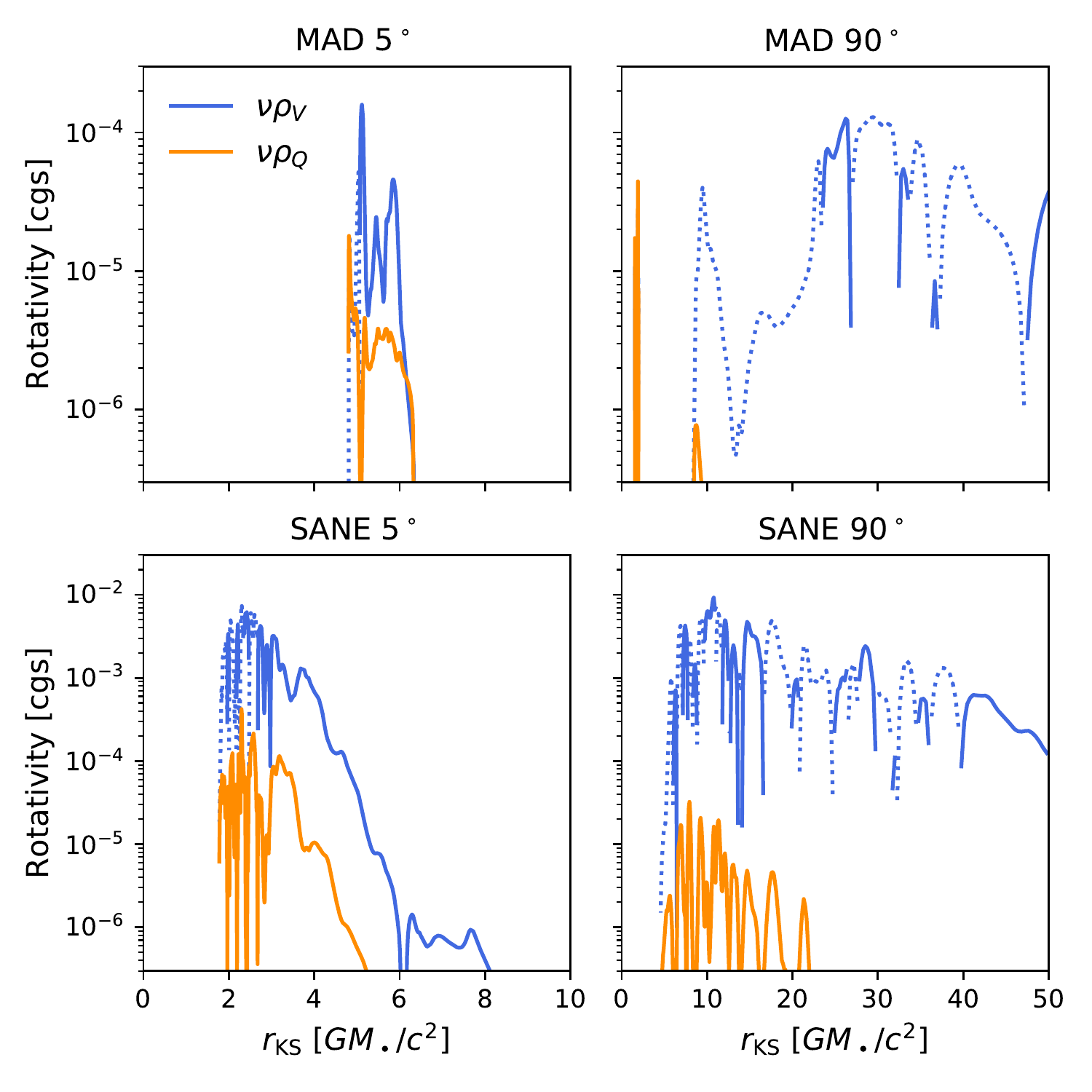}
  \caption{Frame-invariant emissivities (upper panels) and rotativities (lower panels) for representative lines of sight marked with white circles in Figure \ref{fig:faraday_depths}.  We find that both emission and Faraday conversion are enclosed well within the outer integration radius of $20 \ GM_\bullet/c^2$, although Faraday rotation continues out to large radius at $90^\circ$ inclinations.  \label{fig:pixel_traces}}
\end{figure}

Here, we examine representative pixels in each of the models during their final snapshot, marked with white circles in Figure \ref{fig:faraday_depths}.  Along each of these lines of sight, we plot frame-invariant emissivities ($j/\nu^2$; upper panels) and rotativities ($\nu\rho$; lower panels) as a function of radius in Kerr-Schild coordinates, $r_\mathrm{KS}$ \citep[see][for more details]{Moscibrodzka&Gammie2018}.  In the rotativity panels, we omit sections of the geodesic where less than 1 per cent of the final emission along the line of sight has been produced.  This allows us to ignore Faraday rotation and conversion in areas behind the bulk of the emission that do not affect the observables.  This analysis reveals that both the emission and Faraday conversion are enclosed within the outer integration radius of $20 \ GM_\bullet/c^2$ chosen for the radiative transfer.  For $90^\circ$ inclinations, Faraday rotation continues to large radius, as also shown in \citet{Ricarte+2020}, but this does not affect the circular polarization.

\end{document}